\documentclass[11pt,a4paper]{article}
\usepackage[left=2.5cm,right=2.5cm]{geometry}
\usepackage{amsmath}
\usepackage{amsfonts}
\usepackage{amssymb}
\usepackage{graphicx, rotating}
\usepackage{epstopdf}
\usepackage{epsfig}
\usepackage{latexsym}
\usepackage{graphicx}
\usepackage{color}
\usepackage{amsmath,amssymb}
\usepackage{slashed}
\usepackage{hyperref}
\usepackage{booktabs}
\usepackage{multirow}
\usepackage{hyperref}
\hypersetup{colorlinks, citecolor=bluscuro, linkcolor=black, urlcolor=bluscuro}
\definecolor{rossos}{cmyk}{0,1,1,0.55}
\definecolor{bluscuro}{rgb}{0.15, 0.2, .85}
\definecolor{bluchiaro}{cmyk}{1,.3,0.,0.1}

\newcommand{\eq}[1]{Eq.~(\ref{#1})}
\newcommand{\lag}{\mathcal{L}}
\newcommand{\op}{\mathcal{O}}

\newcommand{\nn}{\nonumber}

\newcommand{\dslash}{\!\not\! \partial}
\newcommand{\be}{\begin{equation}}
\newcommand{\ee}{\end{equation}}
\newcommand{\bea}{\begin{eqnarray}}
\newcommand{\eea}{\end{eqnarray}}
\newcommand{\bc}{\begin{center}}
\newcommand{\ec}{\end{center}}

 \def\lra#1{\overset{\text{\scriptsize$\leftrightarrow$}}{#1}}

\begin{document}

\begin{titlepage}
\begin{flushright}
\end{flushright}
\vspace{.3in}

\begin{center}
\vspace{1cm}

{\Large \bf
 Vices and Virtues of Higgs EFTs at Large Energy
}

\vspace{1.2cm}
{\large Anke Biek\"otter$^{a}$, Alexander Knochel$^{a}$, Michael Kr\"amer$^{a}$, Da Liu$^{b,c}$ and Francesco Riva$^{b}$}
\vspace{.8cm}

{\it {$^a$\, Institut f\"ur Theoretische Teilchenphysik und Kosmologie, RWTH Aachen,
Aachen, Germany}}\\

{\it {$^b$\,  Institut de Th\'eorie des Ph\'enom\`enes Physiques, EPFL, 1015 Lausanne, Switzerland}}\\

{\it {$^c$\,State Key Laboratory of Theoretical Physics, Institute of Theoretical Physics, Chinese Academy of Sciences, Beijing, People�s Republic of China}}
  
\begin{abstract}
\medskip
\noindent
We study constraints on new physics from Higgs production at the LHC in the
context of an effective field theory (EFT), focusing on Higgs searches in $HV$
($V=W,Z$) associated production which are particularly sensitive to the
high-energy behavior of certain dimension-6 operators. We show that analyses of
these searches are generally dominated by a kinematic region where the generic
EFT expansion breaks down, and establish under which conditions they can
nevertheless be meaningful. For example, constraints from these searches on the
Wilson coefficients of operators whose effects grow with energy can be
established in scenarios where a particular combination of fermions and the
Higgs are composite and strongly coupled: then, bounds from Higgs physics at
high energy are complementary to LEP1 and competitive with~LEP2.
\end{abstract}

\vspace{.4cm}

\end{center}
\vspace{.8cm}


\end{titlepage}


\section{Motivation}
With the discovery of a Higgs boson\cite{expatlas,expcms}, experiments have finally probed all
sectors of the Standard Model~(SM). The priority is now to measure the
properties of the Higgs particle, and to explore the mechanism of electroweak
symmetry breaking. 
In this paper, we will focus on scenarios beyond the SM (BSM) in which the new
physics modifying the Higgs interactions is heavy\footnote{Scenarios with
additional light non-SM particles are either tightly constrained already or can
be searched for through their modifications of  differential distributions at
small momenta~\cite{Isidori:2013cla,Gonzalez-Alonso:2014rla,Falkowski:2014ffa}.}. Its leading
effects on the SM can then be parametrized through effective operators of
dimension six, suppressed by the scale of new physics $\Lambda$~\cite{Buchmuller:1985jz,Grzadkowski:2010es}.  Some of the
many possible operators which affect Higgs properties can be measured in
Higgs physics only, while others are related to electroweak (EW)
 observables. This is due to the fact that the Higgs scalar
excitation $v+h$ is always associated with the EW symmetry breaking order
parameter~$v$~\cite{Pomarol:2013zra,Gupta:2014rxa,Elias-Miro:2013mua}.
Due to the limited precision of hadron machines, one would think that LHC Higgs measurements are 
unlikely to compete with LEP constraints on this second group of operators.
Nevertheless, the extended energy reach of LHC allows it to access regions
where the effects of some operators are enhanced by powers of
$E/\Lambda$, leading to an increase in sensitivity.

Unlike in on-shell Higgs production by gluon fusion or in Higgs decays, which
occur at $E\sim m_h$, in channels in which the  Higgs is  produced in
association with electroweak gauge bosons, $pp \to hV$, $V=W,Z$, the invariant
mass flowing into the $hVV$ vertex is mainly limited by PDF suppression: these
channels can have enhanced sensitivity to effects growing with energy. We
quantify the extent to which the corresponding cross sections and kinematic
distributions~\cite{Isidori:2013cga,Grinstein:2013vsa} 
can be used to constrain physics beyond the SM. Similar arguments hold, e.g.,  
for analyses of the high-energy tail of the $pp\to Z^*Z^*$ cross section \cite{Gainer:2014hha,Grojean}. 
We find that the naive bounds on the coefficients of dimension-6 operators 
which can be extracted from these measurements are indeed very strong, even with
the limited amount of data available at present. Nevertheless, the $E/\Lambda$ 
enhancement comes at the cost that these measurements are dominated by kinematic
regions where the effective field theory (EFT) expansion has broken down unless 
specific Wilson coefficients are very large. They are therefore meaningless in the 
context of generic EFTs. 

The very motivation for studying EFTs and their Wilson coefficients is that
they allow for a simple parametrization of experimental constraints and an
efficient comparison with large classes of UV theories.  For this reason it is
important to attribute a physical meaning to the Wilson coefficients in terms
of masses, couplings or multiplicities of the BSM sector. It is thus crucial to
understand which classes of theories (if any) can yield and enhancement of Wilson
coefficients contributing to Higgs physics at high energy,
rather than simply assuming that such theories exist. Only then can the
bounds which we extract be thought to carry some information.

With this motivation in mind, we assume that the underlying new physics is under
perturbative control even in the strong coupling limit where it can be thought
of as the effective description of a composite sector. We then integrate out
minimally coupled massive states to match this sector to the SM EFT
description.  We show that one combination of operators, $\mathcal O_W-\mathcal
O_B$, which contributes dominantly to $pp\to hV$ production (but not at tree
level to other tightly constrained observables such as $h\to \gamma \gamma, Z
\gamma$ or EWPTs~\cite{Gupta:2014rxa}) can indeed be enhanced by a strong coupling in the
underlying theory if the Higgs and a particular combination of fermions are
composite and strongly coupled~\cite{Giudice:2007fh}.  In this context, the
bounds derived from $hV$ associated production are surprisingly strong,
complementary to those from LEP1 and competitive with those from  LEP2 (see
also~\cite{Ellis:2014dva,Beneke:2014sba}).  In theories where our bounds are
consistent (and, for instruction, also in theories where they are not), we
compare our results with the bounds from LEP2 measurements of Triple Gauge
Couplings~(TGCs), which receive contributions from the same
operators.

This article is organized as follows: In section~2 we introduce the effective
field theory description of Higgs physics. We choose a basis of operators 
which is particularly well-suited for our needs, as it not only allows a relatively
straightforward interpretation in terms of observables, but can also
be easily matched to relevant models of underlying new physics. We discuss the
connection between $pp\to hV$ and observables in TGCs, and examine the
high-energy behavior of the operators in question. The validity of the
effective field theory description is examined in
section~3 where we consider explicit models of new physics with heavy vector
resonances. In section~4 we analyze the existing data for associated Higgs
production with $W$ or $Z$ bosons in ATLAS. Informed by our discussion of EFT
validity and breakdown, we study which bounds on the coefficients of the 
higher dimensional operators can be established under different assumptions about
the underlying new physics. We conclude in section~5. 
\renewcommand{\arraystretch}{1.4}
\begin{table}[htb]
\begin{center}
\begin{tabular}{|lr|}
\hline
Higgs Physics Only&\\
\hline
\hline
${\cal O}_r=
|H|^2|D^\mu H|^2$ &$\,_1$\\\hline
${\cal O}_{BB}=\frac{{g}^{\prime 2}}{4} |H|^2 B_{\mu\nu}B^{\mu\nu}$&$\,_2$\\\hline
${\cal O}_{WW}=\frac{g^2}{4} |H|^2 W^a_{\mu\nu}W^{a\mu\nu}$&$\,_2$\\\hline
${\cal O}_{GG}=\frac{g_s^2}{4} |H|^2 G_{\mu\nu}^A G^{A\mu\nu}$&$\,_2$\\\hline
${\cal O}_{y_u}  =y_u |H|^2    \bar Q_L \widetilde{H} u_R$ &$\,_1$ \\\hline
${\cal O}_{y_d}   =y_d |H|^2    \bar Q_L Hd_R$ &$\,_1$   \\\hline
${\cal O}_{y_e}   =y_e |H|^2    \bar L_L H e_R$&$\,_1$\\\hline  
${\cal O}_6=\lambda |H|^6$&$\,_1$ \\\hline
   \end{tabular}\hspace{5mm}
\begin{tabular}{|lr|}
\hline
EW and Higgs Physics &\\
\hline
\hline
${\cal O}_W=\frac{ig}{2}\left( H^\dagger  \sigma^a \lra {D^\mu} H \right )D^\nu  W_{\mu \nu}^a$&$\,_2$ \\\hline
${\cal O}_B=\frac{ig'}{2}\left( H^\dagger  \lra {D^\mu} H \right )\partial^\nu  B_{\mu \nu}$&$\,_2$ \\\hline
${\cal O}_{HB}=i g'(D^\mu H)^\dagger(D^\nu H)B_{\mu\nu}$&$\,_2$\\\hline
${\cal O}_T=\frac{1}{2}\left (H^\dagger {\lra{D}_\mu} H\right)^2$&$\,_1$\\\hline
${\cal O}_{Hu} =(i H^\dagger {\lra { D_\mu}} H)( \bar u_R\gamma^\mu u_R)$ &$\,_1$  \\\hline
${\cal O}_{Hd} =(i H^\dagger {\lra { D_\mu}} H)( \bar d_R\gamma^\mu d_R)$   &$\,_1$    \\\hline
${\cal O}_{He} =(i H^\dagger {\lra { D_\mu}} H)( \bar e_R\gamma^\mu e_R)$  &$\,_1$    \\\hline
${\cal O}_{HQ}=(i H^\dagger {\lra { D_\mu}} H)( \bar Q_L\gamma^\mu Q_L)$ &$\,_1$      \\\hline
${\cal O}_{HQ}^{\prime}=(i H^\dagger \sigma^a {\lra { D_\mu}} H)( \bar Q_L\sigma^a\gamma^\mu Q_L)$  &$\,_1$\\\hline
   \end{tabular} 
   \end{center}
     \caption{\it Complete, non-redundant, list of CP-even dimension-6
operators that can potentially contribute to Higgs physics. On the left,
operators that can only affect Higgs
physics~\cite{Elias-Miro:2013mua,Pomarol:2013zra,Gupta:2014rxa}; on the right,
operators already constrained by EW tests.  Flavor indices are summed over
(e.g. $\bar e_R\gamma^\mu e_R$ stands for $\bar e_R\gamma^\mu e_R+\bar
\mu_R\gamma^\mu \mu_R+\bar\tau_R\gamma^\mu \tau_R$). For each operator, we
indicate whether it belongs to class 1 or class 2 in the classification of
\eq{Lops} \cite{Elias-Miro:2013mua}.  Our normalization of the operators
differs from previous literature.}\label{tab:operatorsdim6} 
\end{table}
\section{Dimension-6 Operators in Higgs Physics}\label{sec:dim6}

The lack of direct discovery of BSM physics suggests that, if such physics
exists, it is much heavier than the EW scale and lies beyond the LHC reach. In
this situation new physics (NP) can still leave an indirect imprint in
low-energy observables. This can be efficiently and generically parametrized in
the context of EFTs, corresponding to an expansion
in the SM fields and derivatives over the NP scale $\Lambda$,
\begin{equation}\label{leff}
{\cal L}_{\textrm{eff}}={\cal L}_4+{\cal L}_6+\cdots\, ,
\end{equation}
where ${\cal L}_4$ defines the SM, while ${\cal L}_6$ can be written as a sum of local dimension-6 operators 
\begin{equation}\label{Lops}
{\cal L}_6=\sum_{i_1}  g_*^2\frac{c_{i_1}}{\Lambda^2}{\cal O}_{i_1}+\sum_{i_2}  \frac{c_{i_2}}{\Lambda^2}{\cal O}_{i_2}\, ,
\end{equation}
where we have differentiated between two classes of
operators~\cite{Elias-Miro:2013mua}. The operators ${\cal O}_{i_1}$ involve
extra powers of SM fields, and for this reason must also involve extra
powers of a coupling which we have denoted generically by
$g_*$.\footnote{\label{fnxx}This is most clearly seen by keeping powers of
$\hbar$ explicit in the action $S={\cal L}/\hbar$; then, since a simultaneous
rescaling of $\hbar$ and all the couplings and fields in ${\cal L}$ cannot
modify $S$, the couplings must scale like $\hbar^{-1/2}$ while the fields scale like
$\hbar^{1/2}$: this fixes the coupling-power counting of dimension-6
operators.} The operators ${\cal O}_{i_2}$, on the other hand, involve extra
powers of derivatives and are thus suppressed by the scale $\Lambda$ only. In
this notation, the Wilson coefficients $c_i$ are dimensionless, both in mass and
coupling units. 
If the expansion is valid, $\mathcal L_6$
parametrizes the dominant  contributions of baryon- and lepton-number
preserving NP. Complete sets of dimension-6 operators can be found in
Refs.~\cite{Grzadkowski:2010es,Buchmuller:1985jz,Contino:2013kra,Elias-Miro:2013mua},
expressed in different bases, equivalent up to field redefinitions. 

However, only a few of these operators contribute to Higgs physics, and some of
them to EW physics as well. Hence, a global fit including all operators and all
experiments becomes necessary~\cite{Han:2004az,Pomarol:2013zra,Dumont:2013wma}. This
cumbersome task can be partially avoided (and much physical insight gained)
with an educated choice of basis~\cite{Gupta:2014rxa,Masso}. Ideally, it should 
allow us to identify exactly which operators are tightly constrained by LEP1 experiments
(and can therefore be neglected in LHC physics) and which ones could still
provide measurable deviations from the SM.

The most appropriate basis for our purposes, which describes the relevant
contributions to Higgs physics, shares part of the advantages described in
Refs.~\cite{Gupta:2014rxa,Masso} but can also be quickly matched to specific UV
models, is shown in Table~\ref{tab:operatorsdim6}. We restrict this discussion
to CP-even operators at the leading order in the Minimal-Flavour-Violation
hypothesis~\cite{D'Ambrosio:2002ex}, but both assumptions can be easily
relaxed. The bounds we derive can be quickly translated into other bases
using operator identities such as
\begin{align}\label{silhvsus}
{\cal O}_B&={\cal O}_{HB}+ {\cal O}_{BB}+{\cal O}_{WB}\,,\\
{\cal O}_W&={\cal O}_{HW}+ {\cal O}_{WW}+{\cal O}_{WB}\,,\nn
\end{align}
where 
\begin{equation}
{\cal O}_{HW}=i g(D^\mu H)^\dagger\sigma^a(D^\nu H)W^a_{\mu\nu}\,,\quad 
{\cal O}_{WB}=\frac{gg'}{4} (H^\dagger \sigma^a H) W^a_{\mu\nu}B^{\mu\nu}\,,
\end{equation}
and field redefinitions proportional to the equations of motion (EOM), \begin{align}\label{EOMequalities}
\op_W&=g^2 \left[\frac32 \op_r - \frac14 \sum_{u,d,e}\op_y - \op_6+\frac14 (\op_{HL}^{\prime}+\op_{HQ}^{\prime})\right]\,.\\
\op_B&=g^{\prime 2}\left[-\frac12 \op_T+\frac12 \sum_FY_F\op_{HF}\right]\, .\nn
\end{align}

with $F=\{L_L,e_R,Q_L,u_R,d_R\}$, $Y_F$ the hypercharge, and
\begin{equation}
\op_{HL}\equiv (i H^\dagger {\lra { D_\mu}} H)( \bar L_L\gamma^\mu L_L),\quad 
\op_{HL}^{\prime}\equiv (i H^\dagger \sigma^a {\lra { D_\mu}} H)( \bar L_L\sigma^a\gamma^\mu L_L)\, .
\end{equation}
Indeed, some important features of the dimension-6 Lagrangian 
are best highlighted in other bases. In particular, the
substitution $\op_{WW}\to\op_{HW}$ results in  the strongly interacting light
Higgs (SILH)  basis of
Refs.~\cite{Giudice:2007fh,Contino:2013kra,Elias-Miro:2013mua}, which better
captures the low-energy effects from universal UV theories (where the new
physics only couples  to SM bosons).\footnote{In Ref.~\cite{Giudice:2007fh} also the
operator $\op_r$ was replaced by $\op_H\equiv
\partial_\mu|H|^2\partial^\mu|H|^2/2$ through a field redefinition:
$2\op_r=\sum_{u,d,e}\op_y-2\, \op_H + 4\, \op_6$.} 
The substitution
$\{\op_{W},\op_{B},\op_{HB}\}\to\{\op_{HL},\op_{HL}^{\prime},\op_{WB}\}$ using
Eqs.~(\ref{silhvsus},\ref{EOMequalities}), leads to the basis of
Ref.~\cite{Grzadkowski:2010es} (GIMR in what follows). The GIMR
basis mostly includes vertex corrections\footnote{Although the operator
$\op_{WB}$ does contribute to the $W_3B$ propagator, a combination of the
operator $\op_{WB}$ and operators that modify gauge-boson/fermion vertices, is unconstrained by
LEP1 and is bound only at LEP2 because of its contribution to triple gauge vertices~\cite{Gupta:2014rxa}.},
which makes the connection between operators and observables more straightforward~\cite{Gupta:2014rxa}.
 Furthermore, non-universal theories in which new physics couples to the
different fermions independently, is more easily matched to this basis.

To understand which operators should be included in Higgs physics studies, we
will now briefly discuss which ones are constrained by LEP using the basis 
of Table~\ref{tab:operatorsdim6} (see
Refs.~\cite{Pomarol:2013zra,Gupta:2014rxa} for detailed 
analyses in the bases of Refs.~\cite{Grzadkowski:2010es,Giudice:2007fh}).

The operators on the l.h.s. of Table \ref{tab:operatorsdim6} are 
all of the form $|H|^2\times {\cal L}_{\textrm{SM}}$. In the vacuum ($|H|^2=v^2/2$)
they merely redefine the SM input parameters, and thus at tree level only 
contribute to Higgs physics~\cite{Elias-Miro:2013mua}. 
All these operators 
modify the Higgs vertices and can be constrained by measuring the decay rates 
 $h\to \gamma\gamma,Z\gamma,\bar b b, \bar \tau \tau$, the production
modes $gg\to h$, $VV\to h$, $pp\to \bar t th$ and the trilinear $h^3$ coupling.
At present, however, only the operators $ {\cal O}_{BB},{\cal O}_{WW},{\cal
O}_{GG}$ are constrained tightly enough to justify the EFT expansion (see
section \ref{sec:eftvalidity}).
On the other hand, the operators on the r.h.s. of Table \ref{tab:operatorsdim6}
also affect physics in the vacuum:
$\op_{He},\op_{Hu},\op_{Hd},\op_{HQ},\op_{HQ}^{\prime}$ and the combination
$\op_W+\op_B$ are  constrained by LEP measurements of $Z$-boson
couplings to quark and leptons on the $Z$-pole (for the sake of counting, one
can think of LEP1 as measuring independently the 7 couplings of  $Z$ to $\nu$,
$e_{L,R}$, $u_{L,R}$ and $d_{L,R}$).  These are all tightly
constrained~\cite{Han:2004az,Pomarol:2013zra} and will be neglected in what
follows. The operator ${\cal O}_{HB}$ and the combination $\op_W-\op_B$, on the
other hand, affect in particular Triple Gauge Couplings (TGCs) in addition to
Higgs physics. Measurements of TGCs from diboson production at LEP2 and at the
LHC constrain these operators. However, an analysis in the context of
dimension-6 operators including all existing data is not yet available (see the
discussion in Ref.~\cite{Brooijmans:2014eja}).  For this reason we
include these operators when studying Higgs physics.

Hence, the Wilson coefficients of all operators in Table
\ref{tab:operatorsdim6} are related to some experiment. Consequently,
it is a prediction from ${\cal L}_6$ that any {\it additional} observable which can be
extracted from Higgs physics is already constrained at some level of
precision~\cite{Pomarol:2013zra,Gupta:2014rxa,Masso}. This is true, for example, for
observables contained in channels with $V^*\to Vh$  associated production ($V\equiv W^\pm,Z$), to
which we now direct our attention. Since the s-channel vector is off-shell,
measurements of the differential distributions in these processes can access
regions of momenta where the contribution of some operators is enhanced w.r.t.
their contribution in gluon fusion and Higgs decays. Indeed, amplitudes
such as $q\overline q \rightarrow V_L h$ involving longitudinal massive vector 
bosons will be sensitive to the breaking of gauge invariance communicated by 
the operators of Table~\ref{tab:operatorsdim6}.  In particular, unlike
$\mathcal O_{VV}$, the operators $\mathcal O_{V},\mathcal O_{HV}$ contribute to
Goldstone boson production $q\overline q\rightarrow h G^{\pm,0}$ and
$q\overline q\rightarrow G^{\pm,0}G^{\mp,0}$ and will therefore have the
strongest impact on the high-energy tail of distributions in $V_L h$ and $V_L
V_L$ final states. For $V=W^\pm$, only $\mathcal O_{W}$ and $\mathcal O_{WW}$
contribute to changes in kinematic distributions\footnote{One linear
combination of the other operators on the r.h.s. of Table
\ref{tab:operatorsdim6} can affect these processes through  modifications of
the Higgs branching ratios and wave-function normalization: we will comment on
this in section~\ref{sec:bounds}.} as we can see by observing the squared
partonic matrix element of $f\overline f \rightarrow W^+h$. In the SM,
$|\mathcal M|^2\rightarrow const$ for high energies at tree level, and hence
for $m_H\ll$ TeV, the amplitudes remain perturbatively unitary. {However, in the
presence of $\mathcal O_{W}$ and $\mathcal O_{WW}$, this changes and at large
center-of-mass energy $\sqrt{\hat s}$, 
\begin{eqnarray}
\sum_T \int d \cos\theta |\mathcal{M}_T|^2  &\rightarrow& \frac{4 g^4}{3}  \frac{m_W^2}{\hat{s}}\left(1 + \, (c_{WW} + c_W)\frac{\hat{s}}{\Lambda^2} \right)^2 \,, \nonumber \\
\int d \cos\theta |\mathcal{M}_L|^2  &\rightarrow& \frac{g^4}{6} \left( 1 +  \,c_{W} \frac{\hat{s}}{\Lambda^2} + 4 \, (c_{WW} + c_W)\frac{m_W^2}{\Lambda^2} \right)^2 \,,
\label{crosssection}
\end{eqnarray}
where we have separated the transverse polarizations of the W boson from the
longitudinal one. As expected, the transverse ones are suppressed by a factor
of $m_W^2/ \hat{s}$, which is due to gauge invariance and the fact that the
longitudinal polarization vector is proportional to $p^\mu / m_W$ in the high
energy limit. Note that the expansion makes sense only for $m_{W, h}  \ll
\sqrt{\hat{s}} \lesssim \Lambda$.  From the formulae, we can also see that
unpolarized measurements may mainly constrain the Wilson coefficient $c_W$
because of  its growing energy behavior in the linear part, while for
$\mathcal{O}_{WW}$, due to its tranverse nature (field strength), the leading
term is suppressed by $m_W^2/\Lambda^2$. We observe however that if we could single
out  the transverse polarization of the $W$ boson, we could gain sensitivity
on  $c_{WW}$ (see also \cite{Godbole:2013saa}).  }Similarly, the operators
$\mathcal O_{B}$, $\mathcal O_{HB}$ and $\mathcal O_{BB}$ will enter $Z$
associated production.

\begin{figure}[htb]
\begin{center}
\begin{picture}(230,130)(-13,0)
\put(0,0){\includegraphics[width=7cm]{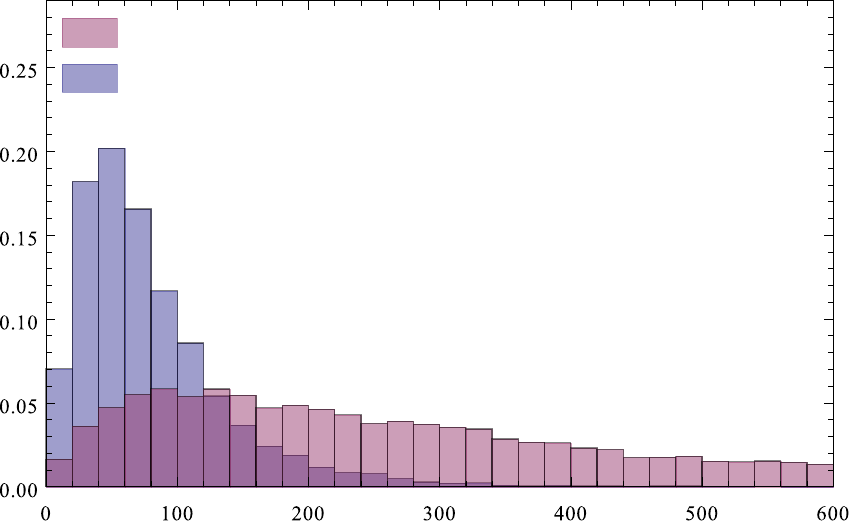} }
\put(-13,40){\scriptsize \rotatebox{90}{$(d\sigma/dp_T)/\sigma$}}
\put(90,-10){\scriptsize $p_T(V)$}
\put(30,112){\scriptsize $c_{W}=0.16 (\Lambda^2\!/\!m_W^2\!), c_{B}=-0.09 (\Lambda^2\!/\!m_W^2\!)$}
\put(30,102){\scriptsize $c_{W}=c_{B}=0$}
\end{picture}
\begin{picture}(220,130)
\put(0,0){\includegraphics[width=7cm]{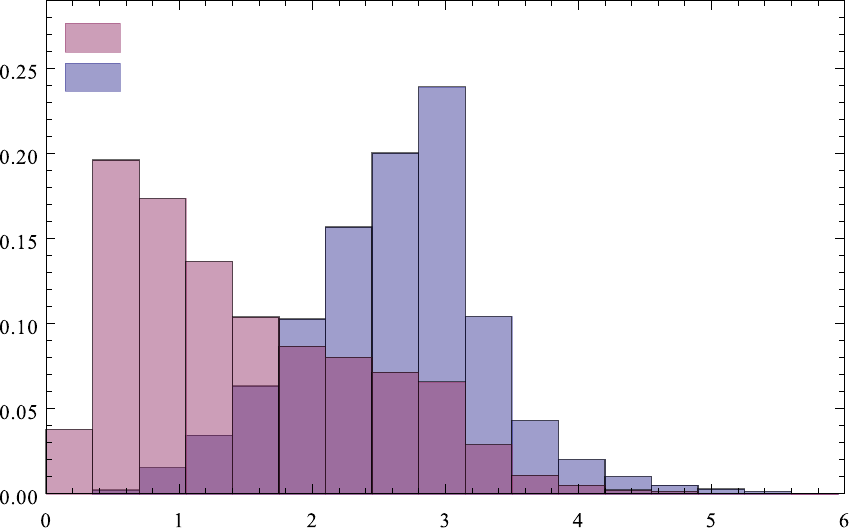} }
\put(-13,35){\scriptsize \rotatebox{90}{$(d\sigma/dR_{b\overline b})/\sigma$}}
\put(95,-10){\scriptsize $\Delta R_{b\overline b}$}
\put(30,112){\scriptsize $c_{W}=0.16 (\Lambda^2\!/\!m_W^2\!), c_{B}=-0.09 (\Lambda^2\!/\!m_W^2\!)$}
\put(30,102){\scriptsize $c_{W}=c_{B}=0$}
\end{picture}
\end{center}
\caption{To illustrate the UV behavior of the operators $\mathcal O_{V}$, these plots
contrast the partonic LO distributions of $p_T(V)$ and $\Delta R(b,\overline
b)$  ($pp\rightarrow ZH$@8TeV) for the SM and SM+$\mathcal O_{V}$ with large Wilson coefficients. \label{pphwpTV}}
\end{figure}
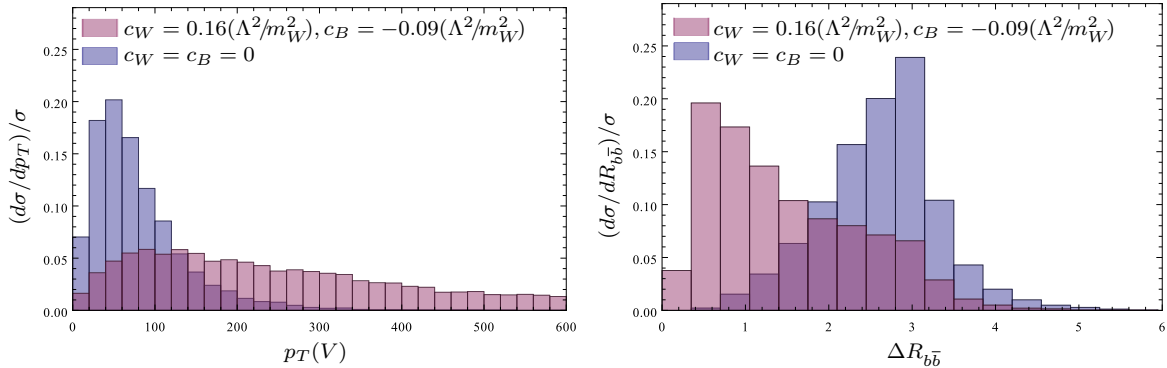

\begin{figure}[tb]
\begin{center}
\begin{picture}(220,210)(-15,0)
\put(0,0){\includegraphics[width=7cm]{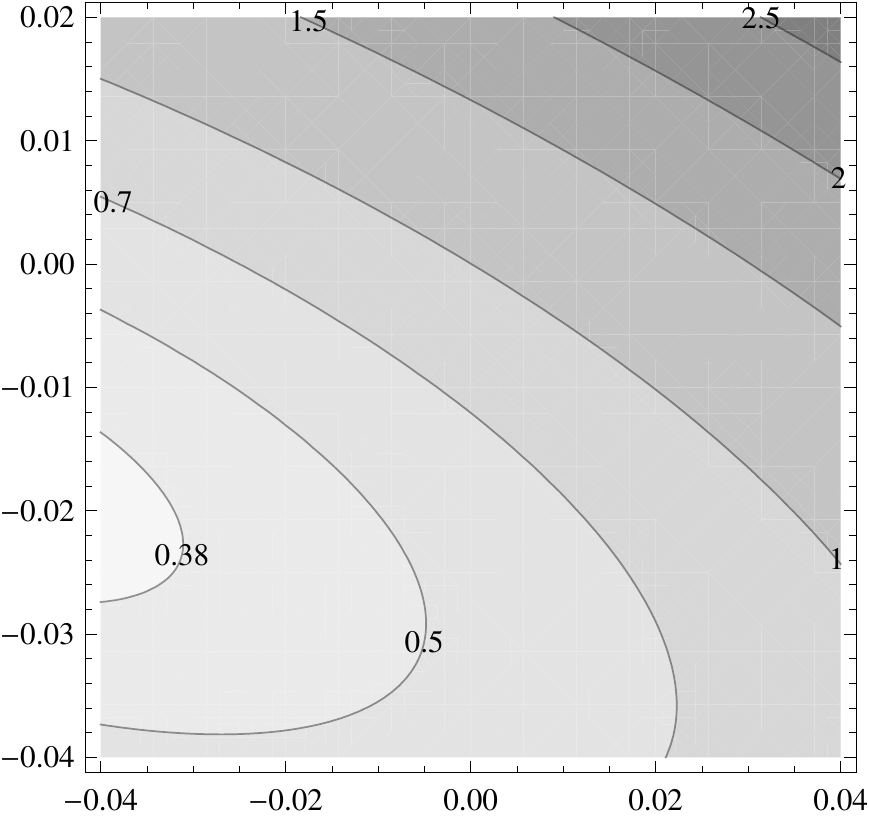}}
\put(75,-15){$c_{WW}(m_W^2/\Lambda^2)$}
\put(-15,70){\rotatebox{90}{$c_{W}(m_W^2/\Lambda^2)$}}
\end{picture}
\begin{picture}(220,210)(-15,0)
\put(0,0){
\includegraphics[width=7cm]{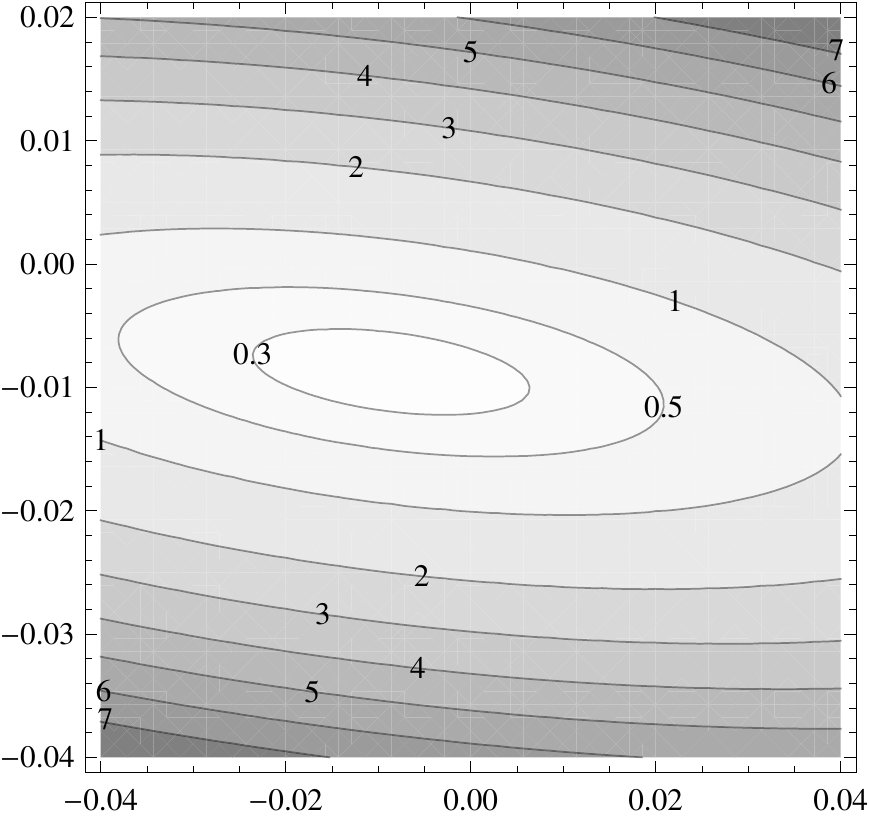}}
\put(75,-15){$c_{WW}(m_W^2/\Lambda^2)$}
\put(-15,70){\rotatebox{90}{$c_{W}(m_W^2/\Lambda^2)$}}
\end{picture}

\end{center}
\caption{The impact of the operators $\mathcal O_{WW}$ and $\mathcal O_{W}$ on
the cross section and kinematics of $pp\rightarrow Wh$ at the LHC8. Shown is $\sigma/\sigma_{SM}$ (LEFT) 
and $\sigma/\sigma_{SM}(p_T>200)$ (RIGHT). The net effect of
$\mathcal O_{WW}$ on the signal strength is subdominant in the region $p_T(W)>200$ GeV. 
We assume that the EFT is valid up to the unitarity cutoff.\label{fig:pphwscan}}
\end{figure}
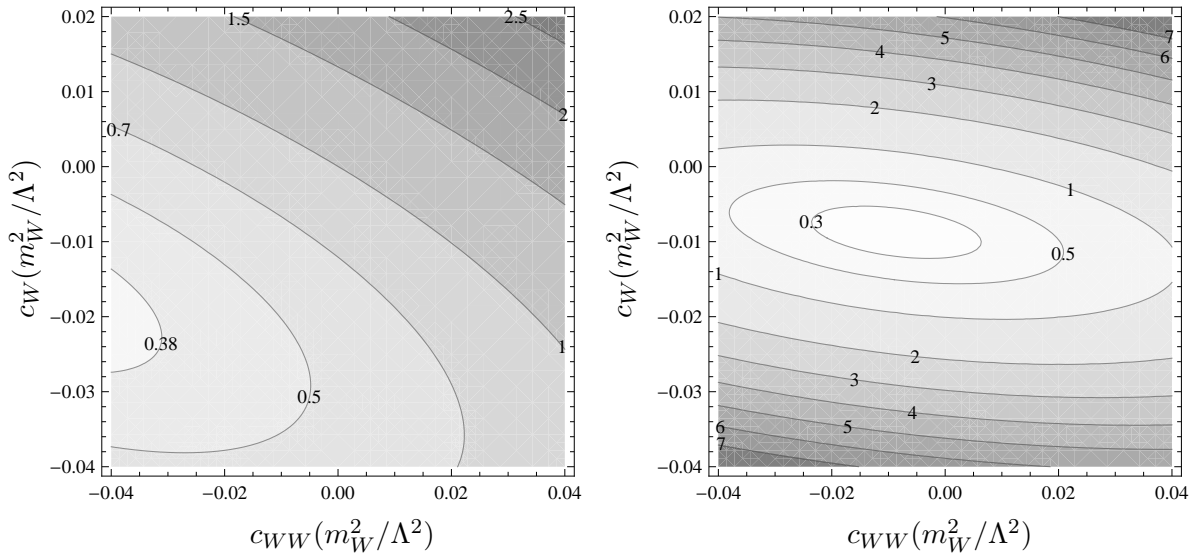

The high-energy behavior of the cross-section, described by \eq{crosssection},
is portrayed in the l.h.s of Fig.~\ref{pphwpTV}, which shows the transverse
momentum $p_T$ distribution for $pp$ collisions.  The high-energy behavior also
impacts the boost distribution of the Higgs in the laboratory reference frame,
which is best captured by the $\Delta R(bb)$ distribution that we show in the
r.h.s. of Fig.~\ref{pphwpTV}.  Notice from \eq{crosssection} that $\mathcal
O_{BB}$ and $\mathcal O_{WW}$ give a smaller relative contribution to
these processes at large $\hat s$, as we also illustrate in Figure
\ref{fig:pphwscan}.  For this reason, we concentrate our discussion on
$\mathcal O_{W}$, $\mathcal O_{B}$ (actually, only the combination $\mathcal
O_{W}-\mathcal O_{B}$ which is unconstrained by LEP1) and $\mathcal O_{HB}$ and comment later on generalizations. 


\section{On the Validity of the EFT  at Large Energy}\label{sec:eftvalidity}

The  EFT of \eq{leff} is an expansion in derivatives and SM fields over powers of
$\Lambda$, defined as the scale where resonant new physics effects should become
visible. Without additional assumptions, the EFT cannot be expected to describe 
processes at energies higher than $\Lambda$ as operators of arbitrary dimension 
are then expected to become equally important, leading to a breakdown of the EFT description. 
In a bottom-up approach (from an IR point of view), $\Lambda$ is not known a priori, 
but is a free parameter which needs to be fixed by experiment. The 
question whether or not the energy at which an experiment is performed lies 
within the validity of the EFT then depends on the sensitivity of the experiment itself. 
For instance, LEP1, working at c.o.m. energy $\sqrt{\hat s}=m_Z$, put bounds 
$\Lambda\gtrsim 1.6$ TeV for operators like the combination $\op_W+\op_B$. The 
sensitivity of the measurement hence fully justifies the EFT expansion in $E/\Lambda$,
making the procedure self-consistent.  As we will see, at least for the Higgs production
data available from the 7 TeV and 8 TeV LHC runs, the situation is less clear.

\begin{figure}
\begin{center}
\begin{picture}(250,140)
\put(0,0){\includegraphics[width=9cm]{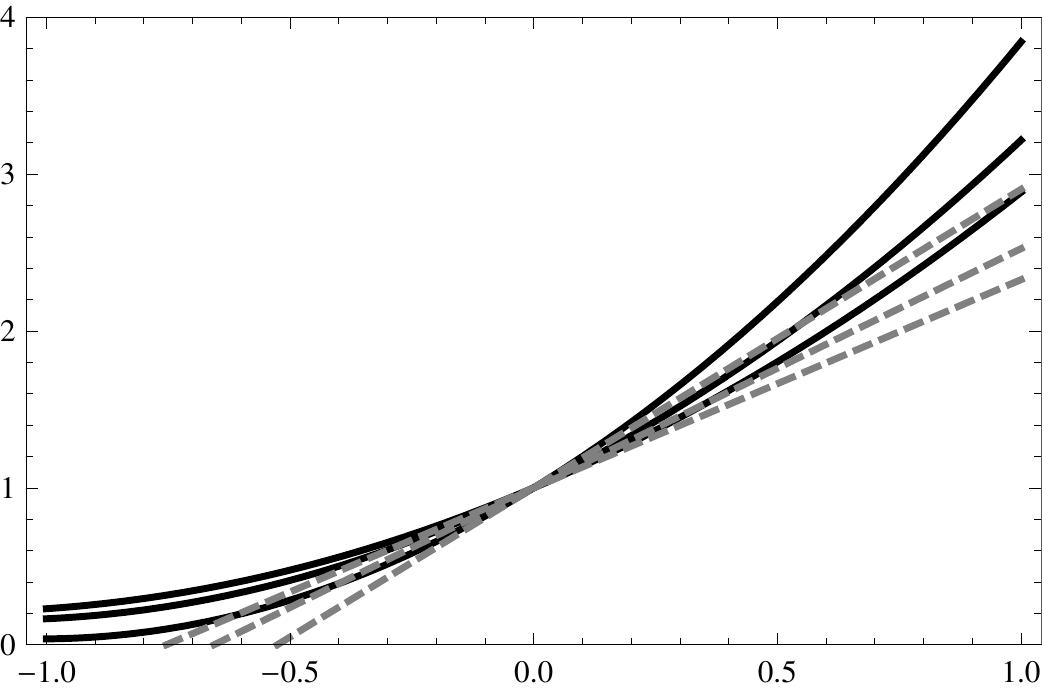}}
\put(100,-10){$c_{W}(m_W^2/\Lambda^2) \times \hat s$}
\put(-15,70){\rotatebox{90}{$\sigma/\sigma_{SM}$}}
\put(220,135){\rotatebox{00}{\small \rotatebox{45}{$1200$}}}
\put(220,115){\rotatebox{00}{\small \rotatebox{45}{$500$}}}
\put(20,25){\rotatebox{00}{\small \rotatebox{0}{$400$}}}
\end{picture}
\end{center}
\caption{The $c_{W}$ dependence of the $u\overline d \rightarrow hW^+$
cross section at different fixed c.o.m. energies $\sqrt{\hat s} =400,500,1200$
as described in the text. All orders of $c_{W}$ are included in the squared
amplitude for the solid lines. The dashed lines represent the linearized signal
strengths.\label{fig:scaling}} \end{figure}

Dimension-6 operators including more derivatives  with respect to an
existing dimension-4 interaction (class $2$ in the classification of \eq{Lops})
are expected to contribute an extra factor of $p^2\sim \hat s$ to the amplitude
compared to the SM, and hence 
\begin{equation}\label{contr2}
\frac{\sigma}{\sigma_{SM}} \sim
(1+c_{i_2} \frac{\hat s}{\Lambda^2})^2\,
\end{equation}
(in reality, this somewhat simplistic view will be complicated by helicity effects). 
For $c_{i_2}\sim O(1)$, the points at which SM
amplitudes are overtaken by EFT effects would typically mark the breakdown of
the expansion in $E/\Lambda$.  This is indeed the case for the operators in
which we are interested. This is illustrated in Fig.~\ref{fig:scaling}, where we show
the $u\overline d \rightarrow hW^+$ cross section in the presence of $\mathcal
O_W$ at fixed center-of-mass energies $\sqrt{\hat s}=400,500,1200$,
and compare the first (linear) term of $\sigma/\sigma_{SM}$ in the $c_W
E^2/\Lambda^2$ expansion with the complete expression.  As expected, at c.o.m.
energies $\sqrt{\hat s}\sim \Lambda/\sqrt{c_W}$, both the linear $O(c_W)$ and
quadratic $O(c_W^2)$ contributions of the dimension-6 Lagrangian to the
cross section become comparable to the SM piece (for $c_W<0$, the linearized
signal strengths vanish already before this point, marking the lower limit of
validity of this approximation). In this case, the question of the validity 
of the EFT is therefore related to the size of the EFT effects relative to the SM.
 
Despite the limitations of generic EFTs, most candidates for underlying models
possess a more complicated structure in terms of different masses, couplings,
and particle multiplicities. Hence, some of the Wilson coefficients of \eq{Lops}
might be parametrically larger (or smaller) for different operators.  In
particular, we have already mentioned in the previous section that under the
assumption that new physics is characterized by a strong coupling $g_*$, the
effective suppression of the operators of class 1 in \eq{Lops}, is
\begin{equation}\label{f}
f\equiv \frac{\Lambda}{g_*}\, 
\end{equation}
in the Lagrangian $\mathcal L \sim \mathcal O_1/f^2+\dots$, as these operators 
imply an expansion in fields which is valid only for small field values: $v/f\ll 1$.
The important point is that $f$ can be parametrically (up to
$1<g_*\lesssim4\pi$ times) smaller than the masses of new
particles $m_*\equiv \Lambda$, which mark the actual
breakdown of the EFT. Crossections which receive contributions growing
with energy
from these operators would make for a perfect probe for new physics. 
Indeed,
\begin{equation}\label{contr1}
\frac{\sigma}{\sigma_{SM}} \sim
(1+c_{i_1} \frac{g^2_*(E)}{g_{SM}^2})^2
\end{equation}
where $g_{SM}$ describes the relevant (weak) SM coupling and we have defined
$g_*(E)\equiv E/f$~\cite{TalkRR}. From \eq{f} we see that  $g_*(E)<g_*$ for the
EFT to be within the real of validity but, contrary to \eq{contr2}, for $g_*\gg
g_{SM}$ the EFT contribution  relative to the SM can now be much bigger than
unity, without exiting the realm of validity of the EFT.  This situation
arises, e.g., for 4-fermion operators and  their contribution
to $2\to2$  scattering growing with energy: under the assumption that $g_*$ is large, it is
possible to study this process at very high energy and obtain very tight
constraints on the Wilson coefficients of these
operators~\cite{Domenech:2012ai}. 

Do these arguments also apply to the operators that enter in Higgs physics? To
answer this question in more detail, we study an explicit model with a spin one vector
resonance $V_\mu^{a}$, triplet under $SU(2)_L$, characterized by a coupling
$g_*\lesssim 4\pi$ that might or might not be strong (similar arguments can be
made with scalar or fermionic heavy states). Beside correctly describing all
weakly coupled UV theories, our simplified model also captures the essence of
strongly coupled scenarios that admit a weakly coupled holographic description
(it is equivalent to a two-site model) in which $V$ is a vector resonance
emerging from the strong sector; the hope is that this description also
qualitatively captures large classes of genuinely strongly coupled
scenarios~\cite{Giudice:2007fh,Contino:2011np}. Heavy vectors have the correct
quantum numbers to mediate interactions between Higgs, gauge boson or fermion
currents, defined as 
\begin{gather}
J^H_\mu\equiv \frac{i}{2} H^\dagger  \lra{D}_\mu H\,,\quad 
J^{Ha}_{\mu }\equiv \frac{i}{2} H^\dagger\sigma^a\lra{D}_\mu H\,,\quad
J^{-H}_{\mu }\equiv \frac{i}{2} H^T \epsilon \lra{D}_\mu H \,,\\
 J^a_\mu \equiv (D^\nu W_{\nu\mu})^a \,,\quad 
J_\mu\equiv  \partial^\nu B_{\nu\mu}\,, \quad
J_\mu^F\equiv\bar F\gamma_\mu F \,,\quad J_\mu^{Fa }\equiv \bar F\gamma_\mu \sigma^a F \,.
\end{gather}
As sketched above, the most favorable scenario where these effects can be
large is the one of strongly coupled theories, from which $V$ emerges as a
composite resonance. Then, any other composite state will couple to $V$ with
strength $g_*$. In particular, this is true for the Higgs field (a light
composite Higgs can arise as a Pseudo Goldstone boson from the strongly
interacting sector~\cite{Kaplan:1983fs}); SM gauge bosons, on the other hand,
are very likely to be (mostly) elementary\footnote{The theoretical difficulties
to realize composite gauge bosons are summarized in Ref.~\cite{Csaki:2011xn}.
Experimental constraints can instead be recast in terms of their contribution
to the operators ${\cal O}_{2B}$ and ${\cal O}_{2W}$ (defined in the text,
below), tightly constrained  by measurements of $e^+e^-\to e^+e^-$ at LEP2
as they correspond to the $Y$ and $W$ parameters~\cite{Barbieri:2004qk}. If the
gauge bosons are elementary, these contributions are proportional to
$(g/g_*)^2$ or $(g^\prime/g_*)^2$ (see Eqs.~(\ref{v1coeffW},\ref{v2coeffB})) and, if
$g_*\gg g$, these measurements provide only mild constraints on the new physics
scale $m_*$.} and in what follows we will assume that they form a separate
(elementary) sector.  Then, the most general renormalizable\footnote{Despite
the appearance, the Lagrangian \eq{lagV} can be associated to a renormalizable
theory based on local gauge invariance, where $V$ acquires its mass via a
Higgs-mechanism \cite{Low:2009di,Pappadopulo:2014qza}.} Lagrangian  describing
universal couplings of $V$ yields 
\be\label{lagV}
{\mathcal L}_\textrm{Universal} =\frac{m_*^4}{g_*^2} \Bigg [\frac{1}{2m_*^2}V^{a\mu}V_{\mu}^{a} + \frac{V^{a\mu}}{m_*} \left( \gamma_H\frac{J^{Ha}_{\mu }}{m_*^3}+ \gamma_V \frac{J^a_\mu}{m_*^3}\right)-\frac{V^{a\mu\nu}V^a_{\mu\nu}}{4m_*^4}+\frac{|D_\mu H|^2}{m_*^4}\Bigg]
\ee
where we have included the Higgs among the strongly coupled states and, in the
spirit of NDA, have written the Lagrangian in a way that keeps the
scaling in powers of masses and couplings manifest~(see footnote~\ref{fnxx}).
The gauge fields, on the other hand, belong to a separate (elementary) sector
and are characterized as usual by
$\lag_{el}=-W^a_{\mu\nu}W^{a\mu\nu}/(4g^2)$ in non-canonical form, and
similarly for $B_\mu,G_\mu^a$. 

In non-universal theories, the BSM sector can also couple to fermions. If the
fermions (or combinations thereof) are also composite, we can write
\be\label{lagF}
{\mathcal L}_F = \sum_F\left(\gamma_F V^{a\mu}  \frac{J_\mu^{Fa}}{g_*^2}+\frac{i}{g_*^2}\bar F\dslash F\right)\, .
\ee
If they are however elementary (or partially composite), the
strong coupling $g_*$  in front of the kinetic term in \eq{lagF} should be
replaced with the appropriate weak coupling.  The coefficients
$\gamma_{H,V,F}\sim O(1)$ quantify the departure from NDA, where they are
expected to be of order unity. For canonically normalized fields, $V\to g_* V$,
$H\to g_* H$, $F\to g_* F$  and $W\to gW$, we obtain the Lagrangian for $V$~\cite{Low:2009di,Pappadopulo:2014qza}
\be\label{lagV2}
{\mathcal L} =\frac{m_*^2}{2}V^{a\mu}V_{\mu}^{a} + V^{a\mu} \left( \gamma_H\, g_*J^{Ha}_{\mu }+\gamma_V \frac{g}{g_*} {J^a_\mu}+\sum_F\gamma_F g_* J_\mu^{Fa} \right)-\frac{1}{4}V^{a\mu\nu }V_{\mu\nu}^a\, .
\ee
Then, integrating out the heavy vector triplets, gives 
\begin{eqnarray}\label{v0befint}
\mathcal L &=&-\frac{1}{2 m_*^2}\left(\gamma_H g_* J^{Ha}_{\mu } + 
\frac{g}{g_*}\gamma_V  J^{a}_{\mu }+\sum_F\gamma_F g_*   J_\mu^{Fa}\right)^2
+\cdots\, \nonumber \\
&=&  c_r \frac{g_*^2 }{m_*^2}{\cal O}_r +c_y\frac{g_*^2}{m_*^2}{\cal O}_y+ c_6\frac{g_*^2 }{m_*^2} {\cal O}_6+\sum_{F=Q,L} c_{HF}^\prime \frac{g_*^2 }{m_*^2} \op_{HF}^\prime \nonumber \\
&+&c_W \frac{{\cal O}_W}{m_*^2}+c_{2W}\frac{{\cal  O}_{2W}}{m_*^2} +\op_{4fermi}+\cdots\,, 
\end{eqnarray}
where we expand in inverse powers of $m_*$, define 
\begin{gather}\label{v1coeffW}
  c_r= - \frac34 \gamma_H^2\,,\quad c_y = \frac{\gamma_H^2}{8}\,,\quad c_6 = \frac{\gamma_H^2}{2} \,,\quad c^\prime_{HF}= -\frac12 \gamma_H \gamma_F  +\frac12 \gamma_V \gamma_F \frac{g^2}{g_*^2} \,, \nonumber\\ 
  c_W= \gamma_H\gamma_V\,,\quad c_{2W}=- \frac{\gamma_V^2}{2}\frac{g^2}{g_*^2}\,. \label{v1aftint}
\end{gather}
and
introduce the operators ${\cal O}_{2B}\equiv (\partial_\mu B^{\mu\nu})^2$ and
${\cal O}_{2W}\equiv(D^\mu W^a_{\mu\nu})^2$ and $\op_{4fermi}$. The latter is
denoting 4-fermion operators irrelevant for our discussion. The dots in
Eqs.~(\ref{v0befint}) denote higher derivative terms resulting
from the momentum expansion in the propagator of $V$.

Similarly we can study the effects of heavy vector singlets under $SU(2)_L$
but, in order to avoid too large violations of custodial symmetry, we preserve
the global $SU(2)_L\times SU(2)_R$ custodial symmetry of the SM and consider
vectors $V^{0\mu},V^{+\mu}$ triplets under $SU(2)_R$~\cite{Low:2009di}:
\be\label{lagB}
\begin{split}
{\mathcal L} =&\frac{ m_*^2}{2} V^{0\mu} V^0_{\mu} + m_*^2 V^{+\mu} V^-_{\mu} + V^{0\mu} \left( \delta_H  \, g_*{J^{H}_{\mu }} + \delta_V\frac{g^\prime}{ g_*} {J_\mu}+\sum_F\delta_F g_* J_\mu^{F}\right)  \\
&+ \frac{1}{\sqrt{2}}  (\delta_H  \, g_*  V^{+\mu}  J^{-H}_\mu + h.c.)-\frac{1}{4} V^{0\mu\nu} V^0_{\mu\nu}-\frac{1}{2} V^{+\mu\nu} V^-_{\mu\nu}\, ,
\end{split}
\ee
In the low energy theory, this yields the coefficients: 
\begin{gather}\label{v2coeffB}
c_r= - \frac34 \delta_H^2\,,\quad c_y = \frac{\delta_H^2}{8}\,,\quad c_6 =  \frac{\delta_H^2}{2} \,,\quad c_{HF}= -\frac12 \delta_H \delta_F  +\frac12 \delta_V \delta_F \frac{g^{\prime 2}}{g_*^2} \,, \nonumber\\ 
  c_B= \delta_H\delta_V\,,\quad c_{2B}=- \frac{\delta_V^2}{2}\frac{g^{\prime2}}{g_*^2}\,.
\end{gather}
We are particulary interested in the coefficients of
the operators $\op_W,\op_B$ and $\op_{HB}$. For the latter, it is clear that it
does not arise at tree level from integrating out minimally coupled vectors, and
its coefficient is therefore suppressed by a loop factor (similar arguments
hold for $\op_{WW},\op_{BB}$). One can thus estimate the coefficient suppressing
the operator $\op_{HB}$ in the Lagrangian \cite{Giudice:2007fh},
\begin{equation}\label{laghb}
\lag_{HB}\equiv\frac{c_{HB}}{\Lambda^2}\op_{HB}
\end{equation}
as
\begin{equation}\label{cutoffHB}
\frac{c_{HB}}{\Lambda^2}\simeq\frac{g^2_*}{16\pi^2 m_*^2}\lesssim 
\frac{1}{m_*^2}
\end{equation}
(up to factors of order-one), where the inequality is saturated for maximally
strongly coupled theories. Hence, these operators should not be trusted at
energies higher than the inverse scale suppressing the operator, i.e.
\eq{laghb} should only be used at energies 
\begin{equation}
E\lesssim \Lambda/\sqrt{c_{HB}}\simeq m_*\,,
\end{equation}
and even then, this is true only for strongly coupled theories.

The operators $\op_W,\op_B$, on the other hand, do arise from vector exchange at
tree-level and (for elementary transverse gauge bosons) are not enhanced by
a strong coupling: they are instead a genuine probe of the new physics resonance
masses, as the coefficient that suppress them scales~as 
\begin{equation}
\frac{c_{W,B}}{\Lambda^2}\simeq
\frac{1}{m_*^2}\,.
\end{equation}
The effects they generate can be extrapolated only to energies $E\lesssim
\Lambda/\sqrt{c_{W,B}}$, as was expected from \eq{contr2}.  Naively, one might
think that in the presence of a large number of nearly-degenerate vectors at
the scale $m_*$, one could obtain an enhancement $c_{W,B}\sim N$ such that the
effective scale that suppresses these operators could be $m_*/\sqrt{N}\ll m_*$
and thus parametrically smaller than the cut-off. However, in
minimally coupled UV scenarios where the heavy vectors are associated with
additional spontaneously broken gauge symmetries, this is not the case. Indeed,
in such a context, the coupling of the Higgs field to these vectors is
characterized by its quantum numbers under these additional local symmetries:
since the SM Higgs field only possesses four degrees of freedom, it cannot
transform under $N$ distinct $SU(2)$ symmetries, but only under a 
linear combination of them.\footnote{In fact, the four d.o.f. of the SM Higgs
doublet can be cast into a $({\bf 2},{\bf 2})$ of $SU(2)_{SM}\times
SU(2)_{BSM}$ and can transform at most under one additional $SU(2)_{BSM}$ gauge
group; this is the model of \eq{lagV2} \cite{Low:2009di,Pappadopulo:2014qza}.}
Furthermore, it is important to recall that the combination $\op_W+\op_B$,
which for universal theories corresponds to the $S$-parameter, is tightly
constrained by LEP1 measurements. On top of this, in most interesting theories,
the coefficients of these operators  are strictly positive $c_{W,B}>0$
\cite{Orgogozo:2012ct}, and consequently the combination $\op_W-\op_B$ which
enters our analysis is already tightly constrained by LEP1. 

Finally, for composite fermions, we see from
Eqs.~(\ref{v1coeffW},\ref{v2coeffB}) that the operators $\op_{HF}^\prime$  and
$\op_{HF}$ are indeed enhanced by the strong coupling: in the Lagrangian
\begin{equation}
\lag_{HF}\equiv \sum g_*^2\frac{c_{HF}}{\Lambda^2}\op_{HF}+g_*^2\frac{c^\prime_{HF}}{\Lambda^2}\op^\prime_{HF}
\end{equation}
the effective coefficient  that multiplies each operator is \begin{equation}
g_*^2 \frac{c^{(\prime)}_{HF}}{\Lambda^2}\simeq \frac{g_*^2}{m_*^2}=\frac{1}{f^2}\,,
\end{equation}
and the discussion of \eq{contr1} applies: in particular, 
there exists a finite energy range $(g/g_*)m_*\lesssim E <
m_*$ where the effect of these operators relative to the SM can be much bigger
than one while the EFT expansion is still valid. That it is indeed still valid
can be seen by looking at the form of dimension-8 operators that arise from
\eq{v0befint}: operators with more derivatives (schematically of the form
$(p^2/m_*^2)\times \op_{HF}$ in momentum space) which contribute to the same
tree-level process, originate from \eq{v0befint} at the next order in the
momentum expansion, and their contribution to \eq{contr1} is
$\sim g_*^2E^4/(g_{SM}^2m_*^4) = g^2_*(E)/g_{SM}^2 ( E^2/m_*^2) =
g^2_*(E)/g_{SM}^2 ( g_*(E)^2/g_*^2)$. This shows that the cutoff is indeed
$m_*$ and not $m_* g_{SM}/g_*$. Notice that this also implies that it is
consistent to keep
contributions of order $(c^{(\prime)}_{HF})^2$, since these are expected to be
much bigger than the contributions from dimension-8 operators to the same
process.

We have thus found a set of operators which can also be studied  in a regime
where their relative contribution to the SM amplitudes is much bigger than one.
How do these operators contribute to $Vh$ associated production? 
As discussed in the previous section, most of the operators $\op^{(\prime)}_{HF}$
are already tightly constrained
by LEP1 as they modify the couplings of the gauge bosons to fermions.
Nevertheless, there is one combination of these operators  which is equivalent to
an overall shift of the Weinberg angle in the gauge-fermion sector and, as
such, cannot be constrained by LEP1~\cite{Gupta:2014rxa}; it can only be
measured as a relative shift between $\theta_W$ as measured in $Z\bar F F$
couplings and  $\theta_W$ as measured in gauge bosons self-couplings or in
Higgs physics. Indeed, this direction is equivalent
to~\cite{Gupta:2014rxa}
\begin{equation}\label{combferm}
\Delta\lag_{F_{tot}}=2\tan^2\theta_W \left(-\mathcal O_T + \sum_FY_F\op_{HF}\right)-\op_{HL}^{\prime}-\op_{HQ}^{\prime}
\end{equation}
which, using \eq{EOMequalities}, can be shown to induce the same effects as~\cite{Gupta:2014rxa,Masso}
\begin{equation}\label{lagftot}
\Delta\lag_{F_{tot}}=\frac{4}{g^2}(\op_B-\op_W)+\op_{\theta_W(Higgs)}\,
\end{equation}
where
\begin{equation}
\op_{\theta_W(Higgs)}=6\,\op_r-\sum_{u,d,e}\op_y - 4\, \op_6\,.
\end{equation}
modifies the Higgs vertices independently of momentum. Indeed it can be easily
seen that \eq{lagftot} contributes only to TGCs (in particular to the parameter
$g_Z^1$~\cite{Hagiwara:1986vm}) or Higgs physics.
Interestingly, from the arguments given above, the contribution to
$\op_W-\op_B$ from the particular direction \eq{combferm}, is enhanced by a
$g_*^2/g^{2}$ factor w.r.t. the naive contribution from universal theories and
provides a motivated context in which the effect of these operators can be
studied at high-energy, as discussed in \eq{contr1}.

This discussion of the breakdown 
of the EFT from a top-down perspective, can be complemented with a bottom-up
approach  (without detailed knowledge of the UV theory) by analyzing
perturbative partial wave unitarity.  An analysis of partial wave unitarity
violation for a several dimension-6 operators has been performed in
\cite{Gounaris:1994cm}. The operators $\mathcal O_{HW}$
and $\mathcal O_{HB}$ imply the constraints
\begin{equation}\label{pertbound}
\hat s \, \lesssim 15.5 \frac{\Lambda^2}{c_{HW}},\, 49 \frac{\Lambda^2}{ c_{HB}}\,.
\end{equation}
Since $\mathcal O_{HW}$ yields by far the strongest unitarity constraint, we
use it as an estimate for the unitarity violation induced by $\mathcal O_W=\mathcal O_{HW}+\mathcal O_{WW}+\mathcal O_{WB}$.
While universal EFTs are far away from saturating this bound, in the case of
\eq{combferm} we will restrict ourselves to values of $g_*$ which satisfy these
constraints.

In summary, we have found that for generic EFTs, extrapolation of the effects
of the dimension-6 operators in a regime where their contribution, relative to
the SM, is bigger than one, is inconsistent with the EFT expansion itself. For
the operators that can contribute to $HV$ associated production, this is true
also in universal theories characterized by a strongly coupled Higgs sector, where
$\op_{HB}$ arises only at loop level, while $\op_W$ and $\op_B$ are suppressed
by the cutoff itself; this situation does not improve in theories with many
vectors.  Nevertheless, in theories in which the particular combination of
fermions reported in \eq{combferm} is composite and part of a strongly coupled
sector, the combination $\op_W-\op_B$ can be enhanced by the strong coupling,
and its effects can be studied also in a regime where its relative contribution
is much bigger than the SM one.\footnote{This is true also in theories where
the gauge bosons are fully composite, but we have mentioned above the theoretical
limitations and the experimental constraints of these theories.} 

\section{Bounds from Existing LHC Higgs Searches}\label{sec:bounds}
As explained in Section~\ref{sec:dim6}, Higgs associated production channels
can probe Higgs interactions at high energy and are particularly sensitive to
BSM interactions like $\op_B$, $\op_{HB}$ etc., whose  contribution strongly increases
with the center-of-mass energy (see \eq{crosssection}). However, in the previous section
we have shown that in generic EFTs, the perturbative expansion breaks
down at large energy when the relative contribution of these operators is
bigger than one. In this case, experiments whose sensitivity is of the order of the
SM contribution or weaker will not be able to put meaningful constraints on
the EFT. We have shown that the same arguments hold in universal theories even
in the strong coupling limit. Within the relatively general framework which we have
considered, namely that of perturbative minimally coupled UV completions, only specific 
scenarios with strongly interacting fermions allow us to to extrapolate the validity 
of the EFT at large energies, and for the operator $\op_W-\op_B$ only.

For this reason, we begin with a study of the $\op_W-\op_B$ direction. It is
important to notice that a study of this combination in isolation (i.e. by
assuming that the coefficients of all other operators are much smaller) makes
sense for a number of reasons. First of all, $\op_W-\op_B$ is orthogonal to
physics from LEP1, meaning that we can ignore LEP1 constraints in our
discussion as well as the other operators that contribute to LEP1
observables: $\{{\cal O}_T, {\cal O}_{Hu},
{\cal O}_{Hd},
{\cal O}_{He}, 
{\cal O}_{HQ},
{\cal O}_{HQ}^{\prime}
\}$ and the combination $\op_W+\op_B$. Also, $\op_W-\op_B$ does not
contribute to the $h\to\gamma  \gamma$ or $h \to Z \gamma$ partial widths or
the $hgg$ coupling, all of which are tightly constrained from LHC
measurements. Thus, beside the analysis of this article, the only constraints
on $\op_W-\op_B$ are from TGC measurements.  Furthermore, from a theoretical
point of view, we have argued that within the class of UV physics we consider,
$c_{HB},c_{WW},c_{BB}$ are very small despite the enhancement of $c_W-c_B$ and,
in case their size would be big enough to be relevant for the experiment, then
the EFT would not be valid. Finally, the contribution of  other operators (such
as $\op_r$ or all remaining operators in table~\ref{tab:operatorsdim6} that
modify the Higgs width) does not grow as fast with energy in the $Vh$
associated production cross section and their impact is negligible if their
coefficients are within the validity of the EFT expansion (see below
\eq{f}).\footnote{\label{footnotexxx}At present, the bounds on these
operators from analyses of Higgs data are not strong enough to justify the EFT
expansion (see
Refs.~\cite{Corbett:2012ja,Pomarol:2013zra,Aad:2013wqa,Dumont:2013wma}). As a
consequence, also effects of order, e.g., $c_r c_W$ in the cross section for
$Vh$ channels can be important: another sign of the breakdown of the
perturbative expansion as these are clearly of the order dimension-8.  .}

We will therefore study constraints on this combination of operators first and
then discuss possible extension. We will repeat the analysis for different choices of
UV realizations: \emph{i)} universal theories (or generic EFTs) where the
scale that suppresses the operator is the cutoff\footnote{While $p_T$ and $\hat
s$ distributions are related, events of any given $p_T$ can in principle have
arbitrarily high energies at large rapidities. Therefore, we will use $\hat s$
cuts to impose cutoffs in what follows. In practice, cutting on $\sum
(m_i^2+p_{iT}^2)^{1/2}$ is a reasonable approximation.} (i.e. Wilson
coefficients of order unity); \emph{ii)} theories  with composite fermions, in
which the Wilson coefficient can be large, implying a large hierarchy between
the scale that suppresses the operator and the cutoff.
 
A naive comparison of the total cross section with measured signal strengths
is inadequate: on the one hand, the effects of $\op_W-\op_B$ are strongest in high $p_T$ bins
which have the lowest SM+Higgs background, while on the other hand it is
precisely those bins which might be probing the breakdown of the EFT.
For this reason, the full differential distribution must be considered; we do
so in this section and discuss the dependence of the bounds obtained on the
choice of cut-off, which we take  consistently into account. Indeed, for
scenario \emph{ii)}, we rely on the strong coupling and use data from energies up to the unitarity bound \eq{pertbound}. 
For scenario \emph{i)}, on the other hand,
we must discard information coming from events  whose energy lies beyond the
region of generic EFT validity for any given value of the coefficients
$(c_W-c_B)/\Lambda^2$.
Since this affects almost exclusively kinematic regions where there is very
little SM background, this cutoff reduces $\chi^2$ and thus yields conservative
exclusions.\footnote{An estimate of the uncertainty in results due to the
breakdown of perturbativity can also be obtained by comparing  the constraints
obtained from linearized signal strengths $\sigma/\sigma_{SM}\approx 1+a\, c_W$
with the full result, which includes contribution of the same order as those of
dimension-8 operators. Using a variable cutoff procedure as we do in this paper 
yields  both  more conservative results for the generic EFT case
as well as allowing the treatment of EFTs with enhanced Wilson coefficients 
in which $O(c^2)$ effects are actually physically meaningful.}

We extract bounds on the coefficients of these operators using present data on
Higgs associated production, and concentrate on the final state with two
$b$-jets, leptons and missing
energy~\cite{TheATLAScollaboration:2013lia,CMS:2013dda}: \begin{eqnarray}
p p &\longrightarrow& Zh; \,\,h\longrightarrow b \overline b, Z \longrightarrow l\overline l \nonumber,\nu \overline \nu \\
p p &\longrightarrow& W^\pm h; \,\,h\longrightarrow b \overline b, W^\pm \longrightarrow \overline l \nu/ l \overline \nu\,.
\end{eqnarray}
We have  implemented the corresponding ATLAS
searches~\cite{TheATLAScollaboration:2013lia}, where data and expected
background and signal events for each $p_T$ bin are reported.  The simulations
are performed using MadGraph
5\cite{Alwall:2014hca}/Pythia\cite{Sjostrand:2006za}/Delphes\cite{deFavereau:2013fsa}
using our FeynRules \cite{Christensen:2008py,Alwall:2014bza} implementation of
the effective theory and the {\it cteq6l1}\cite{Pumplin:2002vw} PDF sets with variable factorization
scale corresponding to the MG5 standard setting.  The analyses of
Ref.~\cite{TheATLAScollaboration:2013lia} use 5 ($2l$ and $1l$) or 3 ($0l$)
different $p_T(V)$ bins separated at $p_T(V)= (0-90,90-120),120-160,160-200,>200$ GeV which are
subject to different additional kinematic cuts. By treating these bins
separately, we gain sensitivity to the shape of the $p_T$ distributions, and in
particular to the high-energy behavior of the EFT.
\begin{figure}
\begin{picture}(220,140)(0,-20)
\put(20,1){\includegraphics[width=0.45\textwidth]{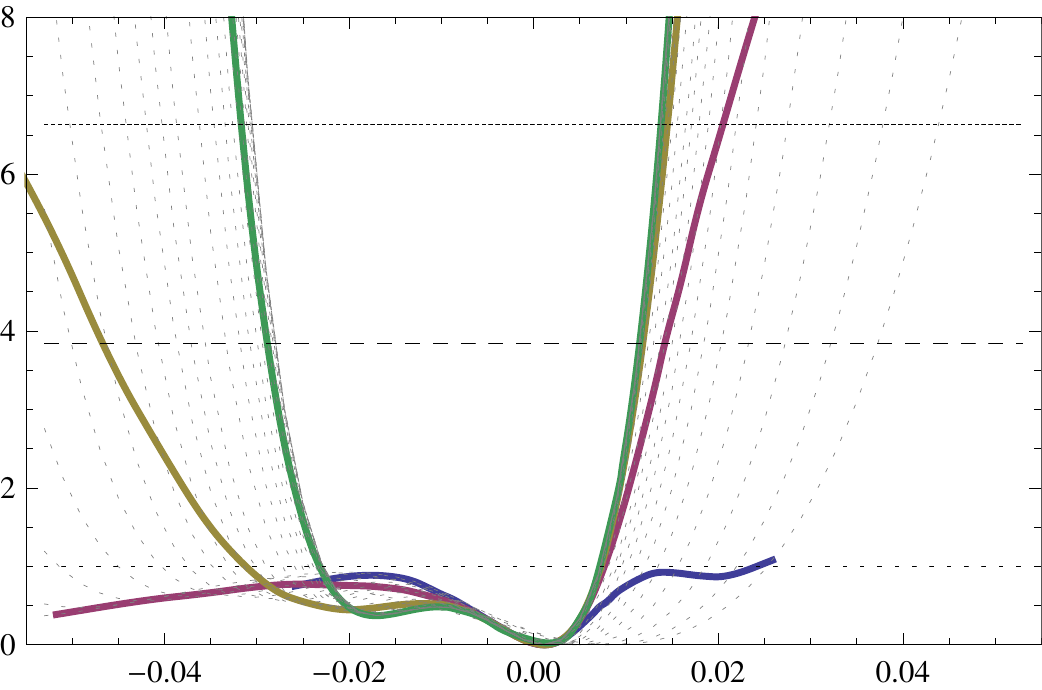}}
\put(5,65){\rotatebox{90}{{\small $\Delta \chi^2$}}}
\put(250,1){\includegraphics[width=0.45\textwidth]{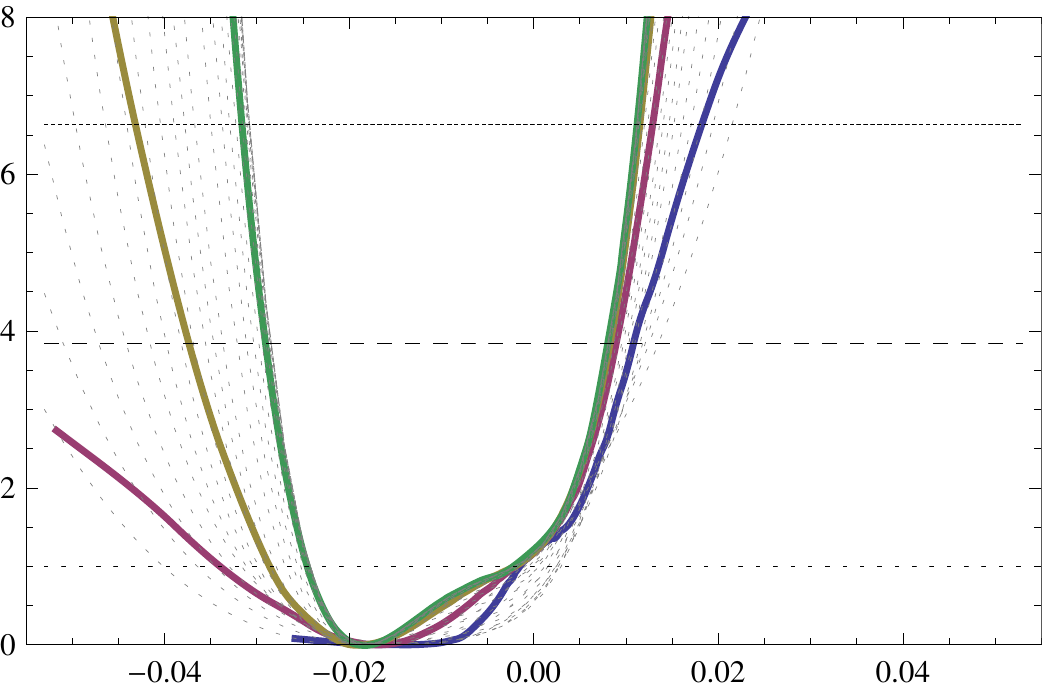}}
\put(90,-10){\small $c_{W}(m_W^2/\Lambda^2)$}
\put(320,-10){\small $c_{W}(m_W^2/\Lambda^2)$}
\put(200,100){$3\sigma$}
\put(200,60){$2\sigma$}
\put(200,15){$1\sigma$}
\end{picture}
\caption{ 
The combined expected (LEFT) and observed (RIGHT) 1-parameter fit $\Delta \chi^2$ contours
in the coefficient $c_{W}(m_W^2/\Lambda^2)=-c_{B}(m_W^2/\Lambda^2)$ from Higgs searches in the $b\overline
b+0l,1l,2l$ final states in ATLAS. We assume all other operators in the basis to be negligible and employ various
UV cutoff prescriptions. The dashed contours are for fixed UV cuts $\sqrt{\hat s} <
500,550,\dots$ GeV, while the solid contours are for parameter-dependent
cutoffs $\hat s< \Lambda^2/c_W $ (blue), $2 \Lambda^2/c_W$ (purple), $4 \Lambda^2/c_W$ (yellow) and $4
\pi\,\Lambda^2/c_W$ (green) inspired by our discussion of UV completions and
perturbativity. 
We assume that the main source of error is systematic, and treat the theoretical errors as nuisances.
\label{fig:unitarityexclusion1d}} 
\end{figure}
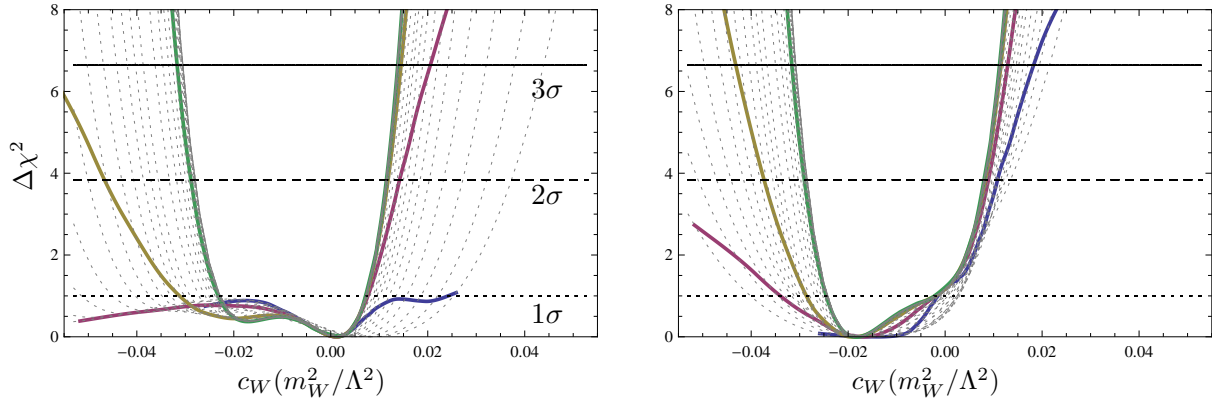

For generic EFTs (case \emph{i)}), $c_{W,B}\sim 1$ and the appropriate cutoff
is the inverse of the scale suppressing the dimension-6 operator. This means
that to every value of the coefficient $1/\Lambda_{}^{2}$ corresponds a
different cutoff $E<\Lambda$. The larger the value of the coefficient, the
smaller the amount of data available to constrain it. As illustrated by
the blue curve in Fig.~\ref{fig:unitarityexclusion1d}, where we plot the
$\Delta\chi^2$ contour for $(c_W-c_B)/\Lambda^2$, the present sensitivity is only
enough to put a constraint on these operators due to an underfluctuation in the 
data, while there is no expected limit as the corresponding $\Delta\chi^2$
never even passes the $2\sigma$ threshold. In the region $c_W=-c_B<0$,
neither the observed nor the expected $\Delta\chi^2$ yield exclusions.  
The same is true for
the $\op_{HB}$ operator with the cutoff suggested by \eq{cutoffHB}.

In case \emph{ii)},  the validity of the EFT is extended to energies
parametrically larger than the inverse of the scale suppressing the dimension-6
operator. The $\Delta\chi^2$  contours obtained with this cut-off correspond to
the purple, yellow and green curves in Fig.~\ref{fig:unitarityexclusion1d}. As the discussion in 
the previous sections indicates, this parametric enhancement depends on the size
of the strong coupling. Since the naive choice for a maximally strong coupling, 
$g_*=4\pi$, violates the partial wave unitarity bound, we limit ourselves
to values of $g_*$ that respect \eq{pertbound} 
(notice however  that for values of $g_*$ as large as $4\pi$ 
the bounds do not change noticeably, see Fig.~\ref{fig:unitarityexclusion1d}).
The corresponding $\Delta \chi^2$ contours are given by the
green curve in Fig.~\ref{fig:unitarityexclusion1d}. We
obtain the following consistent constraint on these operators:
\begin{equation}\label{Higgsbounds}
-0.06\lesssim\frac{c_W-c_B}{\Lambda^2/m_W^2}\lesssim 0.02 \,, \quad 95\%\, \textrm{C.L.}\,.
\end{equation}


\begin{figure}
\begin{center}
\begin{picture}(270,210)
\put(0,0){\includegraphics[width=8.5cm]{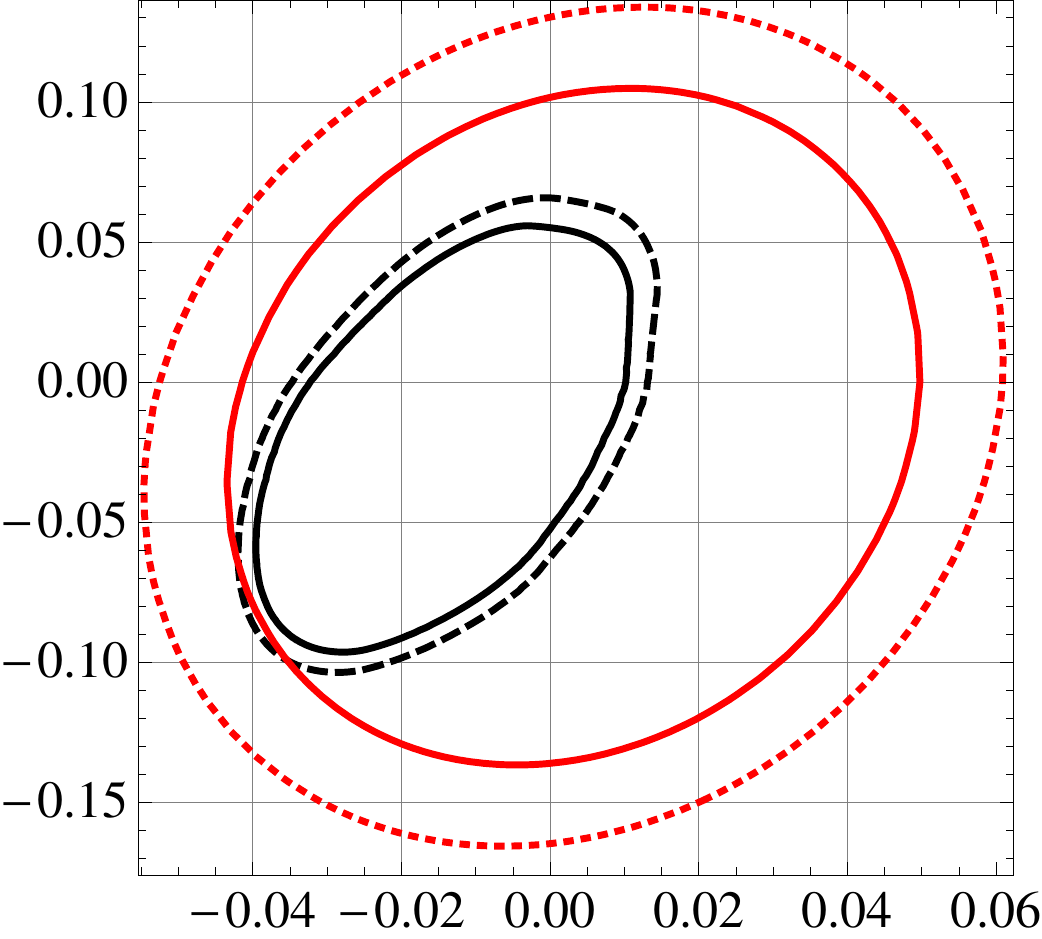}}
\put(80,100){$pp\to Vh$}
\put(175,100){LEP2}
\put(110,-10){$c_{W}(m_W^2/\Lambda^2)$}
\put(-20,80){\rotatebox{90}{$c_{\overline{HB}}(m_W^2/\Lambda^2)$}}
\end{picture}
\end{center}
\caption{The $95CL$ (solid) and $99CL$ (dashed) combined observed limits 
on the coefficients $c_{W}$ and $c_{\overline{HB}}$ (with
$c_B=-c_W$ and all other operators set to zero) from our analysis of Higgs searches in the
$b\overline b+0l,1l,2l$ final states in ATLAS. 
We employ a cut $\sqrt{\hat s}<1.2$ TeV.
We compare the exclusion with
LEP2 limits on TGCs (red contour). \label{fig:unitarityexclusion}} 
\end{figure}
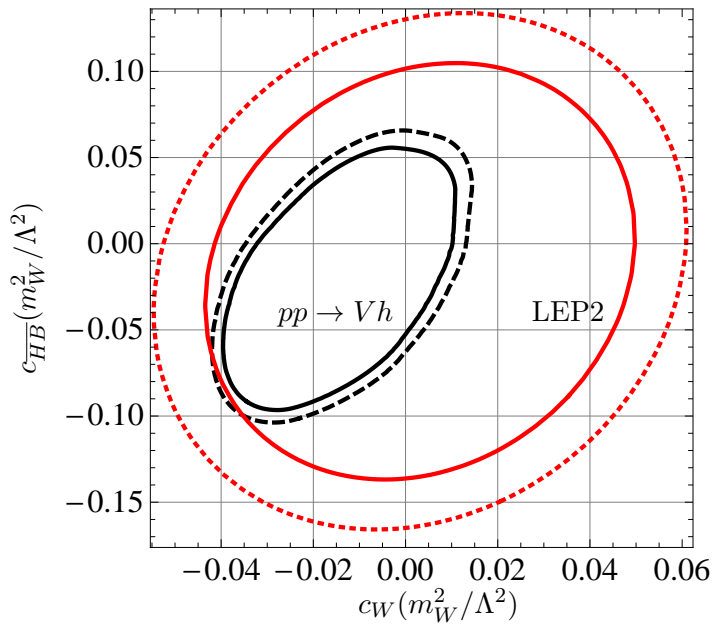

\subsection{Comparison with TGCs}

As explained in section~\ref{sec:dim6}, we want to quantify the added
information of studying channels that probe Higgs physics at high energy. For
this reason we have neglected all operators that are tightly constrained by
either LEP1 or by measurements of $h\to\gamma\gamma,Z\gamma$ and $gg\to h$:
then, only the combinations\footnote{Due a different convention in the
covariant derivative, our definition of $\op_{\overline{HB}}$  differs from
Ref.~\cite{Masso}.}~\cite{Gupta:2014rxa,Masso}
\begin{equation}\op_W-\op_B,\quad \op_{\overline{HB}} \equiv \op_{HB}-\frac12
(\op_{WW}-\op_{BB})
\label{eq:tgcoperators}
\end{equation}
yield contributions to Higgs observables which grow fast with energy and
are not tightly constrained by other experiments (since $\op_{HB}$
contributes to $h\to Z\gamma$, in \eq{eq:tgcoperators} we have cancelled this
contribution by subtracting a piece $(\op_{WW}-\op_{BB})/2$ that contributes
to $h\to Z\gamma$ only~\cite{Gupta:2014rxa,Masso}).  As a matter of fact,
these operators also modify TGCs (measured in $e^+e^-\to W^+ W^-$ scattering at
LEP2), so that it is tempting to compare which experiment gives the strongest
constraints (see also Refs.~\cite{Corbett:2013pja,Corbett:2012ja}).

The sensitivity at LEP2 was high enough to constrain
the Wilson coefficients in TGC measurements within the realm of perturbativity of
generic EFTs. The EFT description at LEP2 is therefore adequately self-consistent.
Nevertheless, there exists at present no analysis of LEP2 data
which consistently includes the effects of all dimension-6 operators (see the
discussion in Ref.~\cite{Brooijmans:2014eja}). A sensible assumption which allows us
to derive bounds on ${\cal O}_{W}-\op_B$ and ${\cal O}_{HB}$ from TGC measurements 
is to limit ourselves to a generic class of theories where the operator ${\cal
O}_{3W}=\frac{\epsilon^{abc}}{3!}W^{a\nu}_\mu W^b_{\nu\rho}W^{c\rho\mu}$ is
small. Under this assumption, the 95\% C.L. bounds from TGCs
are~\cite{LEP2wwzOLD}\footnote{In our basis, the TGC parameters of
Ref.~\cite{Hagiwara:1986vm} are modified as $\delta g_1^Z=c_W/\cos\theta_W^2$
and $\delta\kappa_\gamma=c_{HB}$. As noticed in Ref.~\cite{Brooijmans:2014eja},
under the assumption that $c_{3W}=0$, there is no quantitative difference
between a fit to TGCs in the context of dimension-6 operators (that neglects
terms higher order in the Wilson coefficients) and the fit of the LEP2
collaboration~\cite{LEP2wwzOLD}, which we use in this article.}
\begin{equation}\label{TGCBOUNDS}
 -0.05 \lesssim(\frac{c_{W}-c_B}{2})\frac{m_W^2}{\Lambda^2}\lesssim 0.05\, , \quad\quad -0.12\lesssim c_{\overline{HB}}\,\frac{m_W^2}{\Lambda^2}\lesssim 0.10
\end{equation}
Note that this upper bound on $c_W$ from LEP corresponds to a suppression scale $\gtrsim 350$~GeV,
 larger than relevant LEP2 energies.
 
On the other hand, as discussed above, the constraints from Higgs observables at high-energy that we have derived here are typically beyond the validity of the EFT expansion, but they can make sense for the direction $\op_W-\op_B$, in the case of strongly interacting fermions.
Non-minimally coupled theories could in principle generate tree-level effects for $c_{HB}$,
but it is difficult to argue along the lines of Section~\ref{sec:eftvalidity} 
to say whether the coefficient of these operators can or cannot  be enhanced
with respect to the inverse cutoff.  We assume for completeness that a class of
theories exists where the coefficients of the operator ${\cal O}_{HB}$ can be
very large,  and that the validity of the EFT description can be extrapolated
up to the breakdown of perturbative unitarity. The resulting bounds from
present Higgs data, valid only in this class of theories, are shown in
Fig.~\ref{fig:unitarityexclusion}. We employ a cut $\sqrt{\hat s}<1200$ GeV
corresponding to $\sqrt{4 \pi} m_W/\sqrt{0.05}$, keeping however in mind that 
unlike for $c_{W}-c_{B}$, large values of $c_{HB}$ as they appear in this 
fit are not described in terms
of the UV models presented before.  As mentioned above, we are showing the
direction $\op_{\overline{HB}}$ rather than $\op_{HB}$ only, as
the former gives results that are independent from bounds on $h\to
Z\gamma$~\cite{Gupta:2014rxa,Masso}: in this way the 2D plot shown in
Fig.~\ref{fig:unitarityexclusion} is a genuine comparison between TGCs and
Higgs physics at high energy, and is unaffected by bounds from any other
experiment at present.\footnote{Due to the large coefficients in front of
$c_{WW}$ in \eq{crosssection},  the part $(\op_{WW}-\op_{BB})$ has nevertheless
a sizable impact on he $HV$ channel, although it doesn't grow fast with energy:
for this reason we differentiate between $\op_{HB}$ and
$\op_{\overline{HB}}$.} For this reason, we believe that the plot
of Fig.~\ref{fig:unitarityexclusion} is particularly instructive; furthermore
it quickly allows to differentiate (between vertical and horizontal axes) along
which direction the comparison with TGC makes sense within the context of the
theories described above, and along which one it doesn't.

 \section{Conclusions}
We have investigated constraints on new physics from LHC Higgs searches in an 
EFT context, with an emphasis on channels that are sensitive to high energies 
(in particular Higgs associated production $pp\rightarrow hV$) and can potentially be
very good probes to search for new physics at hadron machines.  In
such Higgsstrahlung processes, the invariant mass is only limited by PDF
suppression (as opposed to on-shell Higgs production, where $\hat s \approx
m_h^2$) and cross sections as well as distributions can be drastically modified
by the presence of dimension-6 operators in the Lagrangian. Indeed, some
operators, unconstrained by LEP1 and by measurements of on-shell Higgs
properties, and only mildly constrained by LEP2 TGC measurements, contribute to
the effective $hVV$ vertex in a manner that grows with energy.

We have concentrated on these operators (namely the combinations $\op_W-\op_B$
and $\op_{\overline{HB}}$ in the basis of Table~\ref{tab:operatorsdim6}) and discussed the
extent to which EFT analyses of LHC Higgs searches can sensibly use the high-energy
tail of distributions to exploit this growth. In
particular, we have shown that in the context of universal theories (where new
physics couples - strongly or weakly - to the SM bosons only) as well as in theories
characterized only by a scale $\Lambda$ and weak couplings, the EFT
expansion is not valid at such large energy. If a consistent analysis, suitable
for universal theories, is performed, then the present data is not accurate
enough to provide any constraints on dimension-6 operators. 

The very essence of using EFTs to parametrize and constrain new physics
BSM is that they can describe large classes of UV scenarios in a simple way and
allow us to quickly reinterpret bounds on the Wilson coefficients as bounds on
masses and couplings of BSM particles. For this reason, rather than simply
assuming the existence of UV scenarios for which the EFT expansion is valid
also at high energy, it is crucial to understand if these scenarios really
exist and which assumptions they require. Therefore, we have explicitly
constructed a class of UV models, characterized by a
strong coupling $g_*$ in addition to the scale $\Lambda$, to study under which
circumstances and for which operators the reach of EFTs can be extended. 
We have found that within this relatively general class of models, one particular
combination of SM fermions needs to be strongly coupled (e.g. as composites emerging
from a strongly coupled sector). This scenario would have been impossible to 
constrain at LEP1, but can be constrained by TGC measurements or through the 
analysis we present here. This is the only concrete scenario we have found 
in which a study of the differential distribution of the $pp\rightarrow hV$ 
channels (and in particular of their high-energy tail), can provide strong 
and consistent constraints on Wilson coefficients. These constraints are 
complementary to LEP1 in the context of fermion compositeness, and are competitive with LEP2.

Furthermore, the indirect limits on anomalous TGCs derived from Higgsstrahlung
using various cutoff prescriptions can serve as a consistency check between
direct searches for new physics and anomalous TGC measurements. 

Finally, the searches outlined in this paper can play an important role in
future high-energy and high-luminosity runs of the LHC, where more precise
measurements of Higgs and gauge boson production rates and kinematics will
compete with direct searches to constrain or discover new physics (see also \cite{Beneke:2014sba}).\\
{\it Note added:} While this paper was in preparation, Ref.~\cite{Ellis:2014dva}
appeared which has some overlap with the present work. While numerical
results agree with the v3 of Ref.~\cite{Ellis:2014dva} where
comparable, our detailed analysis of the EFT breakdown, its impact on LHC Higgs
searches and the interpretation in terms of UV completions are unique to our
work.

\section*{Acknowledgments}
We thank Roberto Contino, Christophe Grojean,  Alex Pomarol, Riccardo Rattazzi,
Dorival G. Netto and Herbi Dreiner for valuable discussions, as well as Simon
Knutzen, Bj\"orn Sarrazin and Klaas Padeken for their assistance with the ROOT
analyses.  We furthermore thank the {\it Centro de Ciencias de Benasque Pedro
Pascual} for their hospitality while parts of this work were in preparation. DL
is  supported by the China Scholarship Foundation and FR  by the Swiss National
Science Foundation, under the Ambizione grant PZ00P2 136932.


\begin{thebibliography}{10}

\bibitem{expatlas} 
  G.~Aad {\it et al.}  [ATLAS Collaboration],
  ``Observation of a new particle in the search for the Standard Model Higgs boson with the ATLAS detector at the LHC,''
  Phys.\ Lett.\ B {\bf 716}, 1 (2012)
  [arXiv:1207.7214 [hep-ex]].
\bibitem{expcms} 
  S.~Chatrchyan {\it et al.}  [CMS Collaboration],
  ``Observation of a new boson at a mass of 125 GeV with the CMS experiment at the LHC,''
  Phys.\ Lett.\ B {\bf 716}, 30 (2012)
  [arXiv:1207.7235 [hep-ex]].
\bibitem{Isidori:2013cla}
  G.~Isidori, A.~V.~Manohar and M.~Trott,
  ``Probing the nature of the Higgs-like Boson via $h \to V \mathcal{F}$ decays,''
  Phys.\ Lett.\ B {\bf 728} (2014) 131
  [arXiv:1305.0663 [hep-ph]].

\bibitem{Gonzalez-Alonso:2014rla}
  M.~Gonzalez-Alonso and G.~Isidori,
  ``The $h \to 4 \ell$ spectrum at low $m_{34}$: Standard Model vs. light New Physics,''
  arXiv:1403.2648 [hep-ph].

\bibitem{Falkowski:2014ffa}
  A.~Falkowski and R.~Vega-Morales,
  ``Exotic Higgs decays in the golden channel,''
  arXiv:1405.1095 [hep-ph].

\bibitem{Buchmuller:1985jz}
  W.~Buchmuller and D.~Wyler,
  ``Effective Lagrangian Analysis of New Interactions and Flavor Conservation,''
  Nucl.\ Phys.\ B {\bf 268} (1986) 621.

\bibitem{Grzadkowski:2010es}
  B.~Grzadkowski, M.~Iskrzynski, M.~Misiak and J.~Rosiek,
  ``Dimension-Six Terms in the Standard Model Lagrangian,''
  JHEP {\bf 1010} (2010) 085
  [arXiv:1008.4884 [hep-ph]].

\bibitem{Pomarol:2013zra}
  A.~Pomarol and F.~Riva,
  ``Towards the Ultimate SM Fit to Close in on Higgs Physics,''
  arXiv:1308.2803 [hep-ph].


\bibitem{Gupta:2014rxa}
  R.~S.~Gupta, A.~Pomarol and F.~Riva,
  ``BSM Primary Effects,''
  arXiv:1405.0181 [hep-ph].

\bibitem{Elias-Miro:2013mua}
  J.~Elias-Miro, J.~R.~Espinosa, E.~Masso and A.~Pomarol,
  ``Higgs windows to new physics through d=6 operators: constraints and one-loop anomalous dimensions,''
  JHEP {\bf 1311} (2013) 066
  [arXiv:1308.1879 [hep-ph]].

\bibitem{Isidori:2013cga}
  G.~Isidori and M.~Trott,
  ``Higgs form factors in Associated Production,''
  arXiv:1307.4051 [hep-ph].
\bibitem{Grinstein:2013vsa}
B.~Grinstein, C.~W.~Murphy and D.~Pirtskhalava,
``Searching for New Physics in the Three-Body Decays of the Higgs-Like Particle,''
arXiv:1305.6938 [hep-ph].

\bibitem{Gainer:2014hha}
  J.~S.~Gainer, J.~Lykken, K.~T.~Matchev, S.~Mrenna and M.~Park,
  ``Beyond Geolocating: Constraining Higher Dimensional Operators in $H \to 4\ell$ with Off-Shell Production and More,''
  arXiv:1403.4951 [hep-ph].


\bibitem{Grojean}
  A.~Azatov, C.~Grojean, A.~Paul and E.~Salvioni,
  ``Taming the off-shell Higgs boson,''
  arXiv:1406.6338 [hep-ph].

\bibitem{Giudice:2007fh}
  G.~F.~Giudice, C.~Grojean, A.~Pomarol and R.~Rattazzi,
  ``The Strongly-Interacting Light Higgs,''
  JHEP {\bf 0706} (2007) 045
  [hep-ph/0703164].
  
\bibitem{Ellis:2014dva}
  J.~Ellis, V.~Sanz and T.~You,
  ``Complete Higgs Sector Constraints on Dimension-6 Operators,''
  arXiv:1404.3667 [hep-ph], v3 to appear soon.

\bibitem{Beneke:2014sba}
  M.~Beneke, D.~Boito and Y.~-M.~Wang,
  arXiv:1406.1361 [hep-ph].

\bibitem{Contino:2013kra}
R.~Contino, M.~Ghezzi, C.~Grojean, M.~Muhlleitner and M.~Spira,
``Effective Lagrangian for a Light Higgs-Like Scalar,''
arXiv:1303.3876 [hep-ph].

\bibitem{Han:2004az}
Z.~Han and W.~Skiba,
``Effective Theory Analysis of Precision Electroweak Data,''
Phys.\ Rev.\ D {\bf 71} (2005) 075009
[hep-ph/0412166].

\bibitem{Dumont:2013wma}
  B.~Dumont, S.~Fichet and G.~von Gersdorff,
  ``A Bayesian view of the Higgs sector with higher dimensional operators,''
  JHEP {\bf 1307} (2013) 065
  [arXiv:1304.3369 [hep-ph]].

\bibitem{Masso}
  E.~Masso,
  ``An Effective Guide to Beyond the Standard Model Physics,''
  arXiv:1406.6376 [hep-ph].

\bibitem{D'Ambrosio:2002ex}
  G.~D'Ambrosio, G.~F.~Giudice, G.~Isidori and A.~Strumia,
  ``Minimal flavor violation: An Effective field theory approach,''
  Nucl.\ Phys.\ B {\bf 645} (2002) 155
  [hep-ph/0207036].

\bibitem{Brooijmans:2014eja}
A. Falkowski, S. Fichet, K. Mohan, F. Riva and V. Sanz contribution in
  ``Les Houches 2013: Physics at TeV Colliders: New Physics Working Group Report,''
  arXiv:1405.1617 [hep-ph].

\bibitem{Godbole:2013saa}
  R.~Godbole, D.~J.~Miller, K.~Mohan and C.~D.~White,
  ``Boosting Higgs CP properties via $VH$ Production at the Large Hadron Collider,''
  Phys.\ Lett.\ B {\bf 730} (2014) 275
  [arXiv:1306.2573 [hep-ph]].

\bibitem{TalkRR}
R.~Rattazzi, talk at \emph{BSM physics opportunities at 100TeV}, CERN 2014

\bibitem{Domenech:2012ai}
  O.~Domenech, A.~Pomarol and J.~Serra,
  ``Probing the SM with Dijets at the LHC,''
  Phys.\ Rev.\ D {\bf 85} (2012) 074030
  [arXiv:1201.6510 [hep-ph]].

\bibitem{Contino:2011np}
  R.~Contino, D.~Marzocca, D.~Pappadopulo and R.~Rattazzi,
  ``On the effect of resonances in composite Higgs phenomenology,''
  JHEP {\bf 1110} (2011) 081
  [arXiv:1109.1570 [hep-ph]].

\bibitem{Kaplan:1983fs}
  D.~B.~Kaplan and H.~Georgi,
  ``SU(2) x U(1) Breaking by Vacuum Misalignment,''
  Phys.\ Lett.\ B {\bf 136} (1984) 183.

\bibitem{Csaki:2011xn}
  C.~Csaki, Y.~Shirman and J.~Terning,
  ``A Seiberg Dual for the MSSM: Partially Composite W and Z,''
  Phys.\ Rev.\ D {\bf 84} (2011) 095011
  [arXiv:1106.3074 [hep-ph]].

\bibitem{Barbieri:2004qk}
R.~Barbieri, A.~Pomarol, R.~Rattazzi and A.~Strumia,
``Electroweak Symmetry Breaking After Lep-1 and Lep-2,''
Nucl.\ Phys.\ B {\bf 703} (2004) 127
[hep-ph/0405040].


\bibitem{Low:2009di}
  I.~Low, R.~Rattazzi and A.~Vichi,
  ``Theoretical Constraints on the Higgs Effective Couplings,''
  JHEP {\bf 1004} (2010) 126
  [arXiv:0907.5413 [hep-ph]].

\bibitem{Pappadopulo:2014qza}
  D.~Pappadopulo, A.~Thamm, R.~Torre and A.~Wulzer,
  ``Heavy Vector Triplets: Bridging Theory and Data,''
  arXiv:1402.4431 [hep-ph].
 
\bibitem{Orgogozo:2012ct}
  A.~Orgogozo and S.~Rychkov,
  ``The S parameter for a Light Composite Higgs: a Dispersion Relation Approach,''
  JHEP {\bf 1306} (2013) 014
  [arXiv:1211.5543 [hep-ph]].
 
\bibitem{Hagiwara:1986vm}
  K.~Hagiwara, R.~D.~Peccei, D.~Zeppenfeld and K.~Hikasa,
  ``Probing the Weak Boson Sector in e+ e- ---> W+ W-,''
  Nucl.\ Phys.\ B {\bf 282} (1987) 253.

\bibitem{Gounaris:1994cm}
  G.~J.~Gounaris, J.~Layssac, J.~E.~Paschalis and F.~M.~Renard,
  ``Unitarity constraints for new physics induced by dim-6 operators,''
  Z.\ Phys.\ C {\bf 66} (1995) 619
  [hep-ph/9409260].

\bibitem{Corbett:2012ja}
  T.~Corbett, O.~J.~P.~Eboli, J.~Gonzalez-Fraile and M.~C.~Gonzalez-Garcia,
  ``Robust Determination of the Higgs Couplings: Power to the Data,''
  Phys.\ Rev.\ D {\bf 87} (2013) 015022
  [arXiv:1211.4580 [hep-ph]].

\bibitem{Aad:2013wqa}
  G.~Aad {\it et al.}  [ATLAS Collaboration],
  ``Measurements of Higgs boson production and couplings in diboson final states with the ATLAS detector at the LHC,''
  Phys.\ Lett.\ B {\bf 726} (2013) 88
  [arXiv:1307.1427 [hep-ex]].

\bibitem{TheATLAScollaboration:2013lia}
  The ATLAS collaboration,
  ``Search for the bb decay of the Standard Model Higgs boson in associated W/ZH production with the ATLAS detector,''
  ATLAS-CONF-2013-079.
\bibitem{CMS:2013dda}
  CMS Collaboration [CMS Collaboration],
  ``Search for the standard model Higgs boson produced in
  association with W or Z bosons, and decaying to bottom
  quarks for LHCp 2013,''
  CMS-PAS-HIG-13-012.

\bibitem{Alwall:2014hca} 
  J.~Alwall, R.~Frederix, S.~Frixione, V.~Hirschi, F.~Maltoni, O.~Mattelaer, H.~-S.~Shao and T.~Stelzer {\it et al.},
  ``The automated computation of tree-level and next-to-leading order differential cross sections, and their matching to parton shower simulations,''
  arXiv:1405.0301 [hep-ph].

\bibitem{Sjostrand:2006za} 
  T.~Sjostrand, S.~Mrenna and P.~Z.~Skands,
  ``PYTHIA 6.4 Physics and Manual,''
  JHEP {\bf 0605}, 026 (2006)
  [hep-ph/0603175].
\bibitem{deFavereau:2013fsa} 
  J.~de Favereau {\it et al.}  [DELPHES 3 Collaboration],
  ``DELPHES 3, A modular framework for fast simulation of a generic collider experiment,''
  JHEP {\bf 1402}, 057 (2014)
  [arXiv:1307.6346 [hep-ex]].
\bibitem{Christensen:2008py} 
  N.~D.~Christensen and C.~Duhr,
  ``FeynRules - Feynman rules made easy,''
  Comput.\ Phys.\ Commun.\  {\bf 180}, 1614 (2009)
  [arXiv:0806.4194 [hep-ph]].
  A.~Alloul, N.~D.~Christensen, C.~Degrande, C.~Duhr and B.~Fuks,
  ``FeynRules 2.0 - A complete toolbox for tree-level phenomenology,''
  Comput.\ Phys.\ Commun.\  {\bf 185}, 2250 (2014)
  [arXiv:1310.1921 [hep-ph]].
\bibitem{Alwall:2014bza} 
  J.~Alwall, C.~Duhr, B.~Fuks, O.~Mattelaer, D.~G.~Ozturk and C.~-H.~Shen,
  ``Computing decay rates for new physics theories with FeynRules and MadGraph5/aMC@NLO,''
  arXiv:1402.1178 [hep-ph].

\bibitem{Pumplin:2002vw}
  J.~Pumplin, D.~R.~Stump, J.~Huston, H.~L.~Lai, P.~M.~Nadolsky and W.~K.~Tung,
  ``New generation of parton distributions with uncertainties from global QCD analysis,''
  JHEP {\bf 0207} (2002) 012
  [hep-ph/0201195].

\bibitem{Corbett:2013pja}
  T.~Corbett, O.~J.~P.~Eboli, J.~Gonzalez-Fraile and M.~C.~Gonzalez-Garcia,
  ``Determining Triple Gauge Boson Couplings from Higgs Data,''
  Phys.\ Rev.\ Lett.\  {\bf 111} (2013) 1,  011801
  [arXiv:1304.1151 [hep-ph]].
  
\bibitem{LEP2wwzOLD}
The LEP collaborations ALEPH, DELPHI, L3, OPAL, and the LEP TGC Working Group, LEPEWWG/TGC/2003-01.


\end{thebibliography}
\end{document}